# The "*Physics of Diagrams*":
# Revealing the scientific basis of graphical representation design

**Sarah Pissierssens**[a], **Jan Claes**[a,*] (ORCiD 0000-0001-7236-4952),
and **Geert Poels**[a] (ORCiD 0000-0001-9247-6150)

[a] Ghent University, Department of Business Informatics and Operations Management,
Tweekerkenstraat 2, 9000 Ghent, Belgium.
* Corresponding author. Address: EB24, Tweekerkenstraat 2, 9000 Ghent, Belgium;
E-mail: jan.claes@ugent.be.

**Abstract.** Data is omnipresent in the modern, digital world and a significant number of people need to make sense of data as part of their everyday social and professional life. Therefore, together with the rise of data, the design of graphical representations has gained importance and attention. Yet, although a large body of procedural knowledge about effective visualization exists, the quality of representations is often reported to be poor, proposedly because these guidelines are scattered, unstructured and sometimes perceived as contradictive. Therefore, this paper describes a literature research addressing these problems. The research resulted in the collection and structuring of 81 guidelines and 34 underlying propositions, as well as in the derivation of 7 foundational principles about graphical representation design, called the "*Physics of Diagrams*", which are illustrated with concrete, practical examples throughout the paper.

## 1 Introduction

The increasingly data-driven and complex problem domains with which today's organizations are confronted, have led to growing attention towards graphical representations. Data from and in information systems is often multivariate and data sets are becoming notoriously big (Ellis & Dix, 2007; Koshman, 2010). When data grows in size and refers to more abstract concepts, it easily becomes incomprehensible (Buagajska, 2003). As a result, people are constructing graphical representations to facilitate data exploration and to ease sense-making (Ellis & Dix, 2007). The increasing interest in graphical representation design in *Information Systems* and *Computer Science* is reflected in the growing research about visualization techniques (Moody, 2009) and quality frameworks for graphical representations (Krogstie et al., 2006; Nelson et al., 2012). Studies frequently refer to research in *Psychology* that shows that textual representations (e.g., natural language, tabular representations) are often insufficient for efficient complex problem solving, and that graphical representation is desirable (Gaissmaier et al., 2012; Regnell et al., 1996).

Well-designed graphical representations have been found to decrease ambiguity (Regnell et al., 1996), to improve understanding (Gaissmaier et al., 2012), to aid processing (Stenning & Oberlander, 1995), to promote completeness of information through inference-making and gap analysis (Cox, 1999; Larkin & Simon, 1987; Novak & Cañas, 2008), to facilitate learning and memorization of knowledge (Novak & Cañas, 2008; Prezler, 2004; Tversky, 2001), to facilitate mental integration of multiple information sources (Sweller et al., 1998), and to support the creation, articulation, communication, and archiving of knowledge (Novak & Cañas, 2008). Moreover, in many fields graphical representations have a pivotal role in the communication of complex information to a wide range of stakeholders. For example, graphical representations have been central from the start to support complex design activities in *Software & System Engineering* (Hjalmarsson & Lind, 2004; Long, 2002; Wagelaar & Van Der Straeten, 2007). In *Business Process Management*, diagrams are the preferred instrument to support activities of process analysis and design (Nelson et al., 2012). In *Environmental & Earth Sciences*, visualization is a critical tool, for example to monitor global environment change (Koshman, 2010; Zahid et al., 1994). In the *Healthcare* domain, information visualization is considered indispensable for the conveyance and understanding of statistical information (Gaissmaier et al., 2012). As a final example, in the *Transportation* industry, visualization techniques ensure that a variety of users are able to extract the right conclusions from safety and risk analysis models (Figueres-Esteban et al., 2015).

Unfortunately, a lot of the communication value of a graphical representation is still determined by the designer (Evitts, 2000). (S)he has to deal with an abundancy of available approaches towards quality, which results in a lack of directedness (Garvin, 1984; Rogers & Scaife, 1998) and in disagreement on what defines the quality of a graphical representation (Moody, 1998; Rockwell & Bajaj, 2005). Second, concrete syntax has been systematically undervalued and put in the back seat. Whereas evaluation techniques for abstract syntax have matured (Berenbach, 2004), a design rationale for concrete syntax has been largely absent (Chen, 2005; Moody, 2009; Rogers & Scaife, 1998). Lastly, the proposed guidelines for graphical representation design are mostly vague, ambiguous, and opinion-based, lacking an underlying structure that explains how properties relate to one another (Nelson et al., 2012).

Therefore, in this paper, we describe a literature study, which resulted in an extensive overview of 81 existing guidelines for graphical representation design, together with the 34 propositions that support these guidelines. The contribution of our work is the linking of the guidelines with the propositions, the structuring of the overview around 12 focus topics, and the derivation of 7 foundational principles for graphical representation design (i.e., the "*Physics of Diagrams*"). It is our belief that this structured overview and the derived principles form a scientific foundation that provides the missing link between the current best-practices and the state-of-the-art academic knowledge about graphical representation design, which should help diagram designers to make informed, optimal design decisions when creating a graphical representation.

This paper proceeds as follows. Section 2 discusses the theoretical background and the problem addressed by this paper, whereas Section 3 explains the research method. The results are presented in Section 4 in the form of an extensive list of 81 guidelines and 34 propositions, grouped in 12 focus topics. This structured overview forms the basis for the derivation of the 7 foundational principles for graphical representation design, presented in Section 5. Finally, Section 6 concludes the paper with a discussion of the results, limitations, and future work.

## 2 Background

A large body of knowledge exists about the definition, quality, and design of graphical representations. Yet, relatively little is known about *how* to create a high-quality representation and *why* certain approaches are preferred over others. This observation is explained in more detail below.

### 2.1 Definition of graphical representation design

A graphical representation (or "diagram") is the product of making abstraction of some of the real-world complexity (Nelson et al., 2012; Rockwell & Bajaj, 2005) by purposefully representing selected information objects and their relationships, in the context of a specific design goal and target audience (Cox, 1999). Diagrammatic representations differ from textual representations on two levels: the encoding and the decoding level (Moody, 2009). To start, in contrast to one-dimensional textual (sentential) representations and to one-to-multi-dimensional tabular representations, diagrams are typically encoded in a two-dimensional solution space (Larkin & Simon, 1987). Secondly, according to dual channel theory (Paivio, 1991), humans decode and process graphical and textual information in separated channels. Consequently, different design principles are required for building graphical and textual representations (Moody, 2009).

In constructing a graphical representation, two aspects need to be considered: the visual notation and the spatial arrangement. The former entails the use of a graphical representation language, which is defined by its concrete syntax (graphical symbols), abstract syntax (vocabulary and grammar), and semantics (the meaning of grammatically correct expressions in the language) (Buagajska, 2003; Moody, 2009). The spatial arrangement reflects the spatial properties of the diagram such as spatial proximity, structural arrangements, sequencing, layering, positioning, reading direction, and spatial density (Buagajska, 2003).

### 2.2 Graphical representation quality

Not every choice, combination and spatial arrangement of graphical symbols will result in a high-quality graphical representation (Austin, 2009). Research has shown that only carefully designed diagrams are advantageous for representing complex information (Tversky et al., 2002). For example, unthoughtful design is shown to negatively impact learning for both high and low prior knowledge learners (Schnotz & Bannert, 2003). Hence, for producing a high-quality diagram, it is important to adhere to proper design rationale.

In the existing literature, many approaches to the concept of quality are taken. Garvin distinguishes three traditional approaches: quality as innate excellence (Tuchman, 1980), quality as requirements conformance (Crosby, 1979), and quality as fitness for use (Juran & Godfrey, 1998). Because of the user-oriented nature of graphical representations (Moody, 2009), we adhere to the latter approach. We thus define quality of graphical representations in terms of their *fit-for-purpose*. This is in line with various other studies that analyze user goals and achievements to determine diagram effectiveness (Casner, 1989; Cox, 1999; Novak & Cañas, 2008).

However, today, a major problem in graphical representation design is the absence of a clear and overarching design goal (Moody, 2009). The absence of such a goal is related to the lack of definition and common understanding of quality in graphical representations (Berenbach, 2004; Rockwell & Bajaj, 2005). What is not concisely and clearly defined, is hard to measure or improve (Ghylin et al., 2007), which may be the reason why the notion of quality in diagrammatic design is still a relatively immature and rapidly evolving concept (Nelson et al., 2012).

Yet, in literature, we find a wide variety of purposes for graphical representations: to facilitate the lookup, comparison, and pattern-marking of data (Roth & Mattis, 1990), to display data distributions (Roth & Mattis, 1990), to show functional relations between elements (Chen, 2005; Roth & Mattis, 1990), to serve as a diagnostic tool (e.g., cognitive mapping (Fiol & Huff, 1992; Siau & Tan, 2005)), to serve as an intuitive language for novices (Fiol & Huff, 1992; Siau & Tan, 2005), to focus attention towards critical information (Fiol & Huff, 1992), to act as an external memory or repository framework and triggering prior knowledge (Fiol & Huff, 1992), to identify bottlenecks, defects and design flows (Sadowska, 2013), to improve, adapt, understand, visualize, automate, and standardize business processes (Sadowska, 2013), to document business activities, data, and knowledge (Rockwell & Bajaj, 2005; Sadowska, 2013), to support software development (Rockwell & Bajaj, 2005; Sadowska, 2013), to improve communication between stakeholders (Sadowska, 2013), to facilitate brainstorming (Sadowska, 2013), to support management initiatives (Nonaka, 2008; Sadowska, 2013), and to coordinate different aspects of real-life task performance (Van Merriënboer & Kester, 2005).

These purposes (some of them domain-independent, others domain-dependent) can be aggregated into three categories: *communication* (with a special ability to focus attention on critical parts), *specification* (to define a concept and show its relations to context elements), and *referencing* (the documentation of information into a repository framework to use for later communication) (De Meyer & Claes, 2018). Further, despite the wide variety, we notice that all the identified purposes for using graphical representations are oriented towards communication and problem solving by humans (Moody, 2009). The challenge is thus to optimize the design of graphical representations such that they are optimally fit for their purpose of supporting human communication.

## 2.3 Towards a scientific basis for graphical representation design

The lack of a scientific basis for diagram design principles (Rockwell & Bajaj, 2005) stimulates the designers to pick and use languages and spatial arrangements based on habits and personal preferences (Moody, 2009; Sweller et al., 1990). On the one hand, this is caused by the abundance of existing language extensions and dialects, which serve different purposes. For example, the Petri Net language has many popular extensions, including Colored Petri Nets and Predicate-Transition Nets (Jensen, 1987, 1991), which has obstructed the exchange of Petri Net models within the field (Billington et al., 2003). On the other hand, not only is it difficult to make an optimal selection from the variety of possible languages and dialects, this unthoughtful design rationale has also led to the inadequate use or adaptation of diagram language variants (Wheildon & Heard, 2005), because the effects of graphic design choices tend to be counterintuitive (Wheildon & Heard, 2005).

A second consequence of this lack of sound graphical representation design principles is that limited progress has been made towards a holistic framework for improving or evaluating graphical representation design (Chen, 2005; Nelson et al., 2012; Rockwell & Bajaj, 2005; Scaife & Rogers, 1996). Although we are still far away from converging towards a common scientific basis for graphical representation design, a first major contribution towards a scientific basis for quality in graphical representation design was made by Moody with his pioneering paper "*The Physics of Notations*" (Moody, 2009). His work provides nine principles for the design of graphical representation languages, which has emerged as a reference theory in the context of graphical representation quality in the *Information Systems* and *Computer Science* disciplines.

Our paper aims to build further on Moody's work, but clearly differs from the "*Physics of Notations*" in that it provides principles for the design of graphical representations (cf. *Fig. 1*). While the design of a graphical representation is an instance-level issue (top-left quadrant on Fig. 1), the design of a graphical representation *language* is a type-level issue (bottom-left quadrant of Fig. 1). The paper of Moody and our paper are thus complementary in nature and together lay the foundation for a scientific basis for quality in graphical representation design, abstracting from the quality of the *content* of the diagrams (right-hand side of Fig. 1).

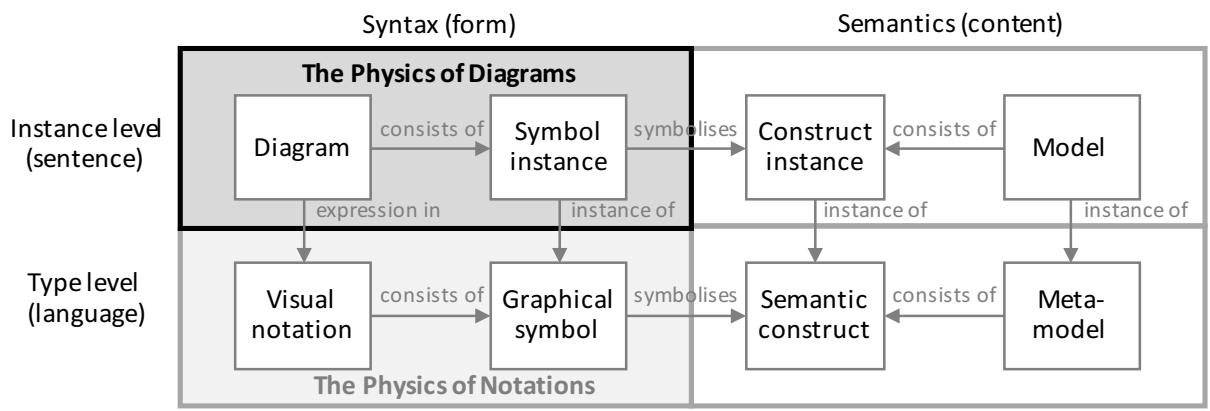

Fig. 1. Whereas Moody's "Physics of Notations" focuses exclusively on the bottom-left hand quadrant, the "Physics of Diagrams" proposed in this paper focuses on the top-left quadrant of the figure. Both artefacts leave the right-hand quadrants out of scope. (adapted from (Moody, 2009))

## 3  Methodology

The research goal is twofold. First, we aim to reveal the tacit knowledge embedded in existing rule-of-thumb guidelines and bring it together with the supporting theoretical knowledge from reference domains. Second, we try to create sensitivity and awareness around the importance of proper design rationale and to point out the scientific character of graphical representation design.

In order to structure the literature study, *Communication Science* was utilized as a guiding grid to comprehensively cover the elements of graphical representations as a communication tool. As established above, the purpose of graphical representations is to facilitate the communication of information between stakeholders. Shannon & Weaver's communication theory (1949) argues that communication consists of four elements: an information source (here: the diagram designer(s)), the message (here: a diagram), a channel that facilitates the communication (here: paper and/or screen), and a destination (here: the diagram user(s)) (Fiske, 1990). Communication theory also distinguishes three phases: encoding, transmitting, and decoding. These elements are represented in Fig. 2.

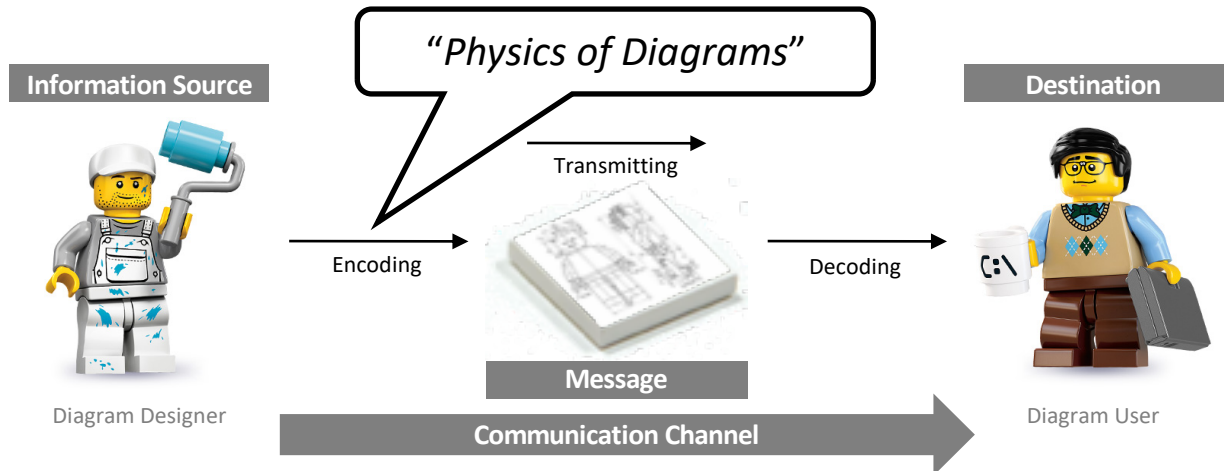

Fig. 2. Shannon & Weaver's communication theory applied to graphical representations situates the "*Physics of Diagrams*" in a communication encoding context.

First, we aim to reveal the tacit knowledge scattered and hidden in existing design guidelines by looking at scientific literature covering all three communication phases. Second, we gathered *propositional* knowledge to scientifically evaluate and underpin the procedural knowledge discovered in the first step. In line with previous research about diagram notations (Rockwell & Bajaj, 2005; Rogers & Scaife, 1998; Zugal, 2013), we investigate propositions from various *cognitive* domains as a promising foundation for the principles to improve the quality of diagrams. Additionally, we searched for relevant literature in fields that produced both propositional and procedural knowledge, such as visual languages in *Computing Science*, *Instructional Science & Educational Psychology*, and *Text Processing* research.

We used the three phases of communication and the procedural-propositional distinction as matrix dimensions to structure the knowledge domains that were addressed by this literature study (cf. *Fig. 3*). Based on this matrix, we conclude that the literature search takes a holistic approach to address the research goal. Indeed, to develop a scientific basis for the design of graphical representations – which is a pure encoding task – relevant input may be found in research domains that address any combination of the three phases of communication.

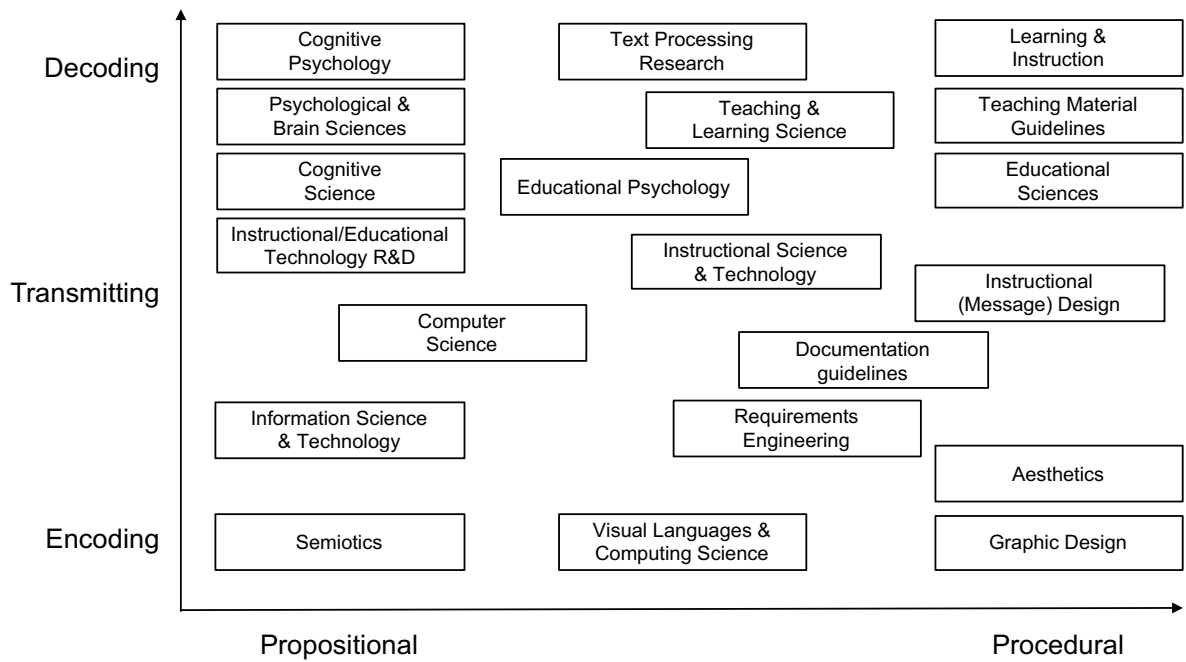

Fig. 3. Matrix representing the searched reference domains according to the three phases of communication theory and distinguishing propositional and procedural literature

Despite the extensive scope of the literature study (spread across a variety of research disciplines, cf. *Fig. 3*), the literature review did not aim for completeness. Therefore, the popular (exhaustive) "*Systematic Literature Review*" method (e.g., Kitchenham et al., 2009; Petticrew & Roberts, 2006) was not followed. Instead, we used a more selective search strategy as in the "*Critical Review*" method (Paré et al., 2015). This search strategy combined several (domain-specific) key phrase searches in different databases (i.e., Google Scholar, Web of Science, and Scopus) with forward and backward snowballing, until the accumulated knowledge extracted from the processed publications reached a stable state. This is similar to the "*structured approach*" described by Webster and Watson (2002).

After revealing the tacit procedural knowledge embedded in prescriptive guidelines and identifying their theoretical grounding in propositional knowledge, the collected knowledge was formally synthesized and integrated. According to their substantive coherence, guidelines and supporting propositions were modularized into representative categories, after which overarching principles were extracted and formulated accordingly. As such, the gap between the propositional and procedural knowledge was bridged through the creation of a well-defined set of scientifically underpinned principles. The aim is to be exhaustive and structurally coherent in the formulation of the principles, to minimize room for confusion or misinterpretation.

## 4   An overview of discovered guidelines and propositions

This section presents an overview of the discovered guidelines and propositions that are deemed relevant for graphical representation design. They are grouped around 12 focus topics, which form the basis for the principles discussed in Section 5. At the end of each group, a summarizing table presents the related propositions and guidelines, and the sources in which they were found.

## 4.1 Diagram understanding

In Section 2.2, it is argued that most diagrams are used (in)directly for human communication. Hence, it is believed that the focus of diagram design should be on understanding (Mendling & Recker, 2007; Zur Muehlen & Recker, 2008). Different propositions explain why certain diagrams are better understood than others.

Research in *Cognitive Science* has found that the capacity of working memory is limited to, on average, 7 elements (Miller, 1956) (**Proposition 01**). According to more recent work by Cowan (2010), only 3 to 5 elements can be simultaneously handled for information processing (**Proposition 02**). Next, the Cognitive Load Theory (**Proposition 03**) states that the cognitive load of a diagram is caused by both the intrinsic complexity of the material (*intrinsic load*) and the way the material is presented (*extraneous load*) (Sweller et al., 1998). When the cognitive load exceeds the working memory capacity, learning is affected and cognitive processing becomes ineffective (Sweller, 1994). Graphical representations should thus be designed to accommodate these limitations. While intrinsic load by definition cannot be changed by graphical design, extraneous load can be minimized by improving the design (Sweller et al., 1998). Further, external representations of information can decrease the need to store the information in working memory (Larkin & Simon, 1987). This ability to reduce the cognitive load is called *computational offloading* (**Proposition 04**) (Larkin & Simon, 1987).

When content elements are separated either in space or time, learners must invest additional cognitive resources mentally to integrate these elements in order to understand the materials (Chandler & Sweller, 1992). This is called the split-attention effect (**Proposition 05**). In the case that related elements are split over *space*, extra cognitive load is imposed by searching and matching elements. When associated elements are split over *time*, effort from the learner is required to recall existing knowledge and to integrate it with the displayed content. Reducing or eliminating the need for mental integration of split content sources, makes that cognitive resources are freed for other tasks.

Table 1. Propositions (P) about diagram understanding found in literature

| Propositions | Sources |
| --- | --- |
| P01. The magical number seven | (Miller, 1956) |
| P02. Cowan's model of attention and memory | (Cowan, 2010) |
| P03. Cognitive Load Theory (CLT) | (Mayer, 1976; Sweller, 1988; Van Merriënboer & Sweller, 2005) |
| P04. Computational offloading | (Larkin & Simon, 1987) |
| P05. Split-attention effect | (Cierniak et al., 2009; Florax & Ploetzner, 2010; Ginns, 2006; Yeung et al., 1999) |

## 4.2 Graphical minimalism

The amount of information that can be stored and processed simultaneously in working memory is limited (cf. *Proposition 01*). Therefore, it makes sense to minimize the amount of different *types* of symbols used in a single graphical representation. We refer to this as the *graphical minimalism* of a diagram.

This can for example be achieved by *introducing symbol deficit* (stop making a distinction in the representation between different concepts, **Guideline 01**) or by *visual filtering* (the feature to (temporarily) hide information in the representation, **Guideline 02**). More concrete, *Learning & Instruction* guidelines emphasize to resist the temptation to add symbol types extraneous to learning (**Guideline 03**), which is proposed to increase clarity (Schnotz, 2013). Brophy & Good (1986) argue that the higher the number of redundant feature and element types on a diagram, the more difficult learning becomes (**Proposition 06**). Excluding these redundant symbol types, i.e., adhering to the *principle of coherence* (**Proposition 07**) (Mayer & Moreno, 2003), has been found to positively impact information transfer and *elaborative learning* (Van Merriënboer & Kester, 2005). Elaborative learning describes the process of learning concepts with extensive and detailed supporting information (Van Merriënboer & Kester, 2005). Similarly, prescriptive guidelines in *Software Engineering* prompt to *respect diagram determinism* (**Guideline 04**) by only modelling the strictly necessary semantical constructs (**Guideline 05**) in order to ensure message conveyance (Ambler, 2005; Sadowska, 2013) and to enable proper modularization (cf. *Section 4.5*), to reduce semantic complexity (**Guideline 06**) (Zugal, 2013).

Table 2. Propositions (P) and guidelines (G) about graphical minimalism found in literature

| Propositions and Guidelines | Sources |
| --- | --- |
| P06. Redundancy hinders learning | (Brophy & Good, 1986; McNamara et al., 1996; Van Merriënboer & Kester, 2005) |
| P07. Principle of coherence | (Van Merriënboer & Kester, 2005) |
| G01. Introduce symbol deficit | (Moody, 2009) |
| G02. Provide visual filtering | (Ellis & Dix, 2007) |
| G03. Resist the temptation to add embellishments (syntax) | (Ambler, 2005; McNamara et al., 1996) |
| G04. Respect diagram determinism (semantics) | (McNamara et al., 1996) |
| G05. Only model the strictly necessary | (Ambler, 2005) |
| G06. Reduce semantic complexity | (Moody, 2009) |

## 4.3 Graphical complexity

Whereas the graphical minimalism of a diagram refers to the number of symbol *types* used, graphical complexity refers to the number of symbol *instances* per type, which should obviously also be controlled to take the limited capacity of working memory under consideration.

Recent research in the field of *Educational Psychology* has found that learning material that is *perceived* to be more supportive, is able to improve learning by reducing *fear of failure* and, as such, increases the confidence and self-efficacy (Cennamo, 2016). Indeed, higher levels of motivation lead to improved self-monitoring and deeper learning strategies (**Proposition 08**) (Pintrich & Schunk, 2002; Sweller et al., 1998). Therefore, it is important that diagrams are readily and accurately perceived and conceived (i.e., *apprehension* principle, **Proposition 09**). Diagram elements that will not be accurately or easily comprehended, should be left out (**Guideline 07**). For example, indicating the precise width of roads on a map adds unnecessary complexity and should be omitted (Tversky et al., 2002). In short, the message for designers is that less can be more.

Further, guidelines from *Software Engineering & Computer Science* advice to prevent overlap and instance redundancy (**Guideline 08**) (Ambler, 2005). Moreover, the ensemble of instances on the diagram should be kept manageable to facilitate schema acquisition (**Guideline 09**) (Moreno, 2006). Complexity management is not only desirable from an effectiveness and efficiency point of view. Research in *Computer Science* has found that often people prefer single-page diagrams (**Guideline 10**) (Ambler, 2005) and simplified information delivery (Gaissmaier et al., 2012), which is related to the motivation of the diagram readers (cf. *Proposition 08*). On the other hand, multimedia-oriented research in *Instructional Science* has found that integrating multiple self-explanatory diagrams into one non-redundant diagram (**Guideline 11**), promotes information transfer and learning (cf. *Proposition 09*) (Van Merriënboer & Kester, 2005).

Table 3. Propositions (P) and guidelines (G) about graphical complexity management found in literature

| Propositions and Guidelines | Sources |
|---|---|
| P08. Higher levels of motivation lead to improved learning | (Gaissmaier et al., 2012; Moreno, 2006; Pintrich & Schunk, 2002) |
| P09. Principle of apprehension | (Tversky et al., 2002) |
| G07. Less is more | (Gaissmaier et al., 2012) |
| G08. Prevent overlap and instance redundancy | (Ambler, 2005) |
| G09. Restrict the number of instance(s) (groups) to $7 \pm 2$ | (Moreno, 2006) |
| G10. Limit diagrams to one page | (Ambler, 2005) |
| G11. Integrate the different sources of information | (Van Merriënboer & Kester, 2005) |

## 4.4 Visual chunking

As explained above, research in *Psychological Science* has found that the working memory is limited and can only successfully process three to five meaningful items at a time (cf. *Proposition 02*) (Cowan, 2010; Rowlatt, 2008). Literature in the field of *Cognitive Psychology* describes an experiment by Bower (1970), who found that learning is improved when material is organized into categories, facilitating learning through the category interrelationships (**Proposition 10**). This principle is in line with an experiment by Lowe (2003) in the field of *Learning & Instruction*, which showed that relatively clear visual-spatial characteristics on diagrams positively affect information extraction (cf. *Proposition 07*).

These experiments demonstrate the effectiveness of visually segmenting material into meaningful groups to improve learning from diagrams (De Westelinck et al., 2005; Van Merriënboer & Kester, 2005), which is called *chunking* (**Proposition 11**). It is a form of visual, non-hierarchal grouping, also referred to in *Information Systems & Computer Science* as '*horizontal modularization*' (De Meyer & Claes, 2018). The technique of chunking fastens feature recognition, value lookup, and inference making (Larkin & Simon, 1987). Additionally, chunking facilitates structural coherence through avoiding the use of labels to denote connectivity between elements (Larkin & Simon, 1987). Furthermore, the Gestalt Theory (**Proposition 12**) (Wertheimer, 1923) supports the use of visual techniques, such as chunking, to indicate item relatedness in a diagram (Koshman, 2010). The theory argues that the perception of objects in an environment is not defined by the sum of the individual instances, but rather by the total configuration of the elements together. Appendix A presents examples of the visual techniques of the Gestalt Theory.

Practically, chunking can be done by first distinguishing advisably 3 to 5 categories of information (**Guideline 12**). Although research indicates that the optimal number of chunks is around 3 to 5 (cf. *Proposition 02*), the amount of information per unit is not particularly constrained in size and complexity (Sweller, 1988). After the identification of different chunks of information, visual variables and techniques can be used to express category distinction. Following the Gestalt Theory, we distinguish 4 ways to chunk information in a display area: by proximity, similarity, symmetry or common fate (Koffka, 2013).

The law of proximity in the Gestalt Theory states that the human mind perceives objects that are close together as a group of related items. Chunking by *spatial proximity* (**Guideline 13**) is often recommended in *Graphical Design* guides (Bertin, 1981), and is frequently applied in the field *Information Systems & Computer Science* (Ware, 2005). To chunk by *similarity* (**Guideline 14**), the designer can make use of visual variables such as color, shape, and texture to indicate element relatedness without physically connecting them with arrows (Koshman, 2010). In *Data Science*, color is often used to highlight datasets that share mutual attributes to facilitate correct data selection (Koshman, 2010; Ware, 2004). Visual chunking is also possible by positioning objects *symmetrically* around a focal point (**Guideline 15**). The Gestalt Theory states that the human mind connects symmetrical elements even though they are not physically connected. If the objects are similar, the chances are higher they will be recognized as related symmetrical elements. By organizing objects *on the same 'trend line' or with the same orientation or direction* (**Guideline 16**), they are more easily perceived as a group as whole. For example, by giving elements the same orientation, their similarity can be communicated. Visual variables such as color and shape can be used to further enhance category distinction (**Guideline 17**). Ware (2005) promotes the use of a combination of contour, color, motion or texture to segment the display space into regions. More guidelines for the use of visual variables to increase expressiveness are discussed in Section 4.10.

Table 4. Propositions (P) and guidelines (G) about visual chunking found in literature

| Propositions and Guidelines | Sources |
|---|---|
| P10. Material organized in categories facilitates learning | (De Westelinck et al., 2005; Gerjets et al., 2004; Van Merriënboer & Kester, 2005) |
| P11. Principle of chunking | (Cox, 1996; Van Merriënboer & Kester, 2005) |
| P12. Gestalt Theory | (Koshman, 2010) |
| G12. Group diagram information into 3 to 5 units | (Sweller, 1988) |
| G13. Chunk by spatial proximity | (Bertin, 1981; Koshman, 2010; Rockwell & Bajaj, 2005) |
| G14. Chunk by similarity | (Koshman, 2010; Ware, 2004) |
| G15. Chunk through symmetry | (Koshman, 2010) |
| G16. Chunk by common fate | (Koshman, 2010) |
| G17. Use visual variables to enhance category distinction | (Ware, 2005) |

## 4.5 Hierarchical chunking

After visually chunking the information into sub-components (cf. *Section 4.4*), hierarchy can be added to the diagram by making the relations between diagram elements explicit. This latter step is called *hierarchical chunking* or *vertical modularization* (**Proposition 13**) (De Meyer & Claes, 2018). The technique is proposed by both research in *Instructional Science* (Gerjets et al., 2004) and in *Information Systems & Computer Science* (Ambler, 2005; Regnell et al., 1996).

The first way to modularize, is to take a top-down approach, where the most inclusive and often abstract concepts are decomposed into their lower-level, often more specific, sub-components (Novak & Cañas, 2008). While decomposing the diagram, it is important to *gradually* increase element refinement (**Guideline 18**) so that the element relationships are obvious (Van Merriënboer & Sweller, 2005). A technique to measure this gradualism, is to count the number of modules 'calling' each specific module (i.e., fan-in) and the number of modules that 'are called' by each module (i.e., fan-out) (Gruhn & Laue, 2006). When a module with both high fan-in and high fan-out is encountered, this signals that the diagram should be further improved (Gruhn & Laue, 2006). As for *nesting depth,* it is important to balance between the size of each module and the amount of modules (**Guideline 19**) (Miller, 1956). The number of modules between the top-level and any bottom-level module is ideally no more than 7 (cf. *Proposition 01*) (Miller, 1956). Respecting this explicit complexity limit is important to maintain *processing fluency* of thoughts (Kuhn & Stahl, 2003; Reber et al., 2004) and motivation of the reader (Cennamo, 2016; Pintrich & Schunk, 2002; Reber et al., 1998).

The second approach to modularize information is to start from the concrete concepts and work your way up to the abstract ones by the process of *summarization* (Berenbach, 2004; Moody, 2009). Summarization allows to *fade* (hide or omit) certain chunks of information (**Proposition 14**) and provide an overview without cluttering the diagram with details (Regnell et al., 1996). The modularized diagram structure is foldable, permitting to show the diagram at different levels of refinement (Berenbach, 2004). In literature, the strategy of fading has been found to improve understanding and to decrease unproductive searching and learning activities (Paas et al., 2004). Through summarization via *recursive decomposition* (DeMarco, 2002), smaller modules are fitted into larger modules, allowing to turn lower level information into black boxes (**Guideline 20**). The use of black boxes improves communication because it guides attention towards the critical parts of the representation and frees up working memory capacity (Paas et al., 2004). Additionally, it avoids duplication of work by allowing reuse of schema parts (Gruhn & Laue, 2006; Regnell et al., 1996), which according to *schema theory* (**Proposition 15**), increases user performance (Paas et al., 2004). Lastly, it is important to ensure strong diagram cohesion (**Guideline 21**) by carefully balancing the level of abstraction and fragmentation. A coherent diagram ideally keeps related elements together but fosters abstract thinking (Zugal, 2013).

Table 5. Propositions (P) and guidelines (G) about hierarchical chunking found in literature

| Propositions and Guidelines | Sources |
|---|---|
| P13. Principle of hierarchical modularization | (De Westelinck et al., 2005; Mautone & Mayer, 2007; Novak & Cañas, 2008) |
| P14. Principle of fading | (Paas et al., 2004) |
| P15. Schema theory | (Sweller et al., 1998) |
| G18. Gradually increase element refinement | (Van Merriënboer & Sweller, 2005) |
| G19. Keep the depth of decomposition manageable | (Cox, 1999) |
| G20. Introduce black boxes to increase diagram complexity | (DeMarco, 2002) |
| G21. Ensure strong diagram cohesion | (Zugal, 2013) |

## 4.6 Direction of information

Continuity is an important factor in the understanding of graphical representations (Ware et al., 2002). The Gestalt *law of good continuation* (**Proposition 16**) describes the human ability to integrate or connect separated components in a graphical representation, under the condition that the flow direction is clear and continuous (Field et al., 1993). An illustration of the law of good continuation can be found in Appendix A. In order for diagrams to benefit from this human cognitive ability of continuation, related graphical elements should be aligned along a common diagram path (**Guideline 22**), which is preferably continuous and smooth (e.g., no zigzagging or brisk, unexpected turns) (Field et al., 1993).

Further, according to research in *Information Systems & Computer Science*, straight lines, either vertically or horizontally, are easier for a user to follow (Ambler, 2005). Diagram elements should be modelled along orthogonal lines (**Guideline 23**), as if they were put on the unit grid of the diagram (Ambler, 2005; Irani & Ware, 2003). Likewise, literature in *Information Systems & Computer Science* (Ambler, 2005; Becker et al., 2000; Long, 2002) recommends to maintain consistent information flow direction within and across diagrams (**Guideline 24**). As in the Western world the general reading direction is from left to right, direction of information should ideally be from left to right as well. The guideline of consistency in flow direction also applies for the direction of arrows, text labels, and shapes of container elements (in case the latter indicate direction) (Evitts, 2000).

Table 6. Propositions (P) and guidelines (G) about direction of information found in literature

| Propositions and Guidelines | Sources |
| --- | --- |
| P16. Gestalt law of good continuation | (Field et al., 1993) |
| G22. Align graph elements on continuous and smooth paths | (Field et al., 1993) |
| G23. Draw edges along orthogonal vertices | (Purchase, 2002; Schuette & Rotthowe, 1998) |
| G24. Maintain consistent flow direction | (Purchase, 2002) |

## 4.7 Internal and external linkage

Mental integration of information is a critical antecedent of learning, but it puts high demands on working memory (**Proposition 17**) (Chandler & Sweller, 1992). Research in *Experimental Psychology* finds that the better integrated learning material is, the greater long-term memory and resistance against interference (Houston, 1965; Saltz, 1971; Sweller, 1988). The integration of information is twofold: the diagram information should be *internally* linked (connecting sub-components within the material), but preferably also linked to relevant *external* elements (connecting to the external context).

When material is separated over either display space, time, or information medium, learners must divide their cognitive capacity over the different elements (Ginns, 2006), and a so-called *split-attention effect* can arise (cf. *Proposition 05*). However, empirical testing found that the split-attention effect does not apply for high-interactive information (Cierniak et al., 2009). Further, the repleteness effect (**Proposition 18**) states that diagrams that use visual notations that were previously acquired by the user (the user knows the symbols) or that are semantically transparent (the user can easily infer the meaning of symbols, **Proposition 19**), improve learning and can lead to beneficial computational offloading effects (De Westelinck et al., 2005).

Concerning internal linkage, drilling down into low-level details of the model too soon in the first phases of graphical representation design, may lead to losing track of the primary goal and to compromise the overarching logic and structure of the diagram (Berenbach, 2004). Therefore, the first design effort should be holistic and focused at the complete diagram breadth (**Guideline 25**) (Berenbach, 2004).

Furthermore, for low-complexity information, empirical evidence encourages to make the relationships between elements explicit (**Guideline 26**), arguing that the *scattering* of elements on a diagram leads to decreased structural coherence (cf. *Proposition 07*) (Caillies et al., 2002; Rockwell & Bajaj, 2005). By providing a well internally-integrated graphical representation, working memory capacity can be freed from integrative processing aspects and instead used for task performance (Carlson et al., 1990; Ginns, 2006).

Similarly, research in *Experimental Psychology* recommends to integrate multiple mutually referring information sources (e.g., text with complementary graphical representations) as much as possible, instead of fragmenting them for aesthetical reasons (Sweller et al., 1990). On the other hand, for high-complexity information, evidence in *Information Systems & Computer Science* states that *weak coupling*, i.e., minimizing the number of connections between instances, can be beneficial (**Proposition 20**) (Zugal, 2013). Therefore, it is recommended to use text segmentation and labelling instead of spatial integration or linking (**Guideline 27**) (Cierniak et al., 2009; Florax & Ploetzner, 2010; Ginns, 2006). An illustration can be found in Appendix B.

Concerning external linkage, research in *Cognitive Psychology* has found that, even though users may have prior knowledge that can help them understand new material (Gemino & Wand, 2003), they are often not able to recognize it themselves (Paris & Lindauer, 1976; Spires & Donley, 1998). *Prior knowledge activation* is therefore recommended (**Guideline 28**) in order to retrieve relevant existing knowledge from long-term memory (Ormrod, 1999). Next, providing an overview of how new material relates to the user's existing knowledge and other problem domains (i.e., contextualizing, **Guideline 29**), has been found to positively impact learning and information transfer (Ormrod, 1999; Spence, 2007; Sweller et al., 1998). In some cases, our concept knowledge is even depending upon the context (e.g., a cup may become a vase when there are flowers in it) (Ormrod, 1999).

Additionally, in *System Information & Computer Science*, it is recommended that developers use modeling languages that (are similar to what) they already know (**Guideline 30**), to facilitate understanding (cf. *Proposition 18*) (De Westelinck et al., 2005). Further, various authors promote *reuse of patterns* in conceptual modeling design (**Guideline 31**) to ease diagram acquisition (Maiden & Sutcliffe, 1992; Nelson et al., 2012; Snoeck & Poels, 2000). The benefit in analogical reuse is that every time a pattern is applied successfully, an accumulation of learning processes lead to internal strengthening and automation of the pattern (Carlson et al., 1990; Van Merriënboer & Kester, 2005). Research in *Cognitive Psychology* has found that the mind seems to store incoming information in the form of previously mastered cognitive structures (Saltz, 1971). Learning information elements that are situated at end points of a dimension (e.g., ice-cold versus hot) are easier to remember (Pollio, 1968). Therefore, it may be beneficial to use extremes of well-known dimensions in the use of word choice and visual variables (**Guideline 32**).

Finally, annotations can provide contextual or additional background info for novices (**Guideline 33**). Annotations may increase control over presented information and readability of the diagram (Boyle & Encarnacion, 1998). For example, Ambler describes the use of *summary notes* (or so called *legends*) in UML diagrams (Ambler, 2005), used to describe the purpose and the broader context of the diagrams.

Table 7. Propositions (P) and guidelines (G) internal and external linkage found in literature

| Propositions and Guidelines | Sources |
|---|---|
| P17. Mental integration | (Rockwell & Bajaj, 2005) |
| P18. Repleteness effect | (De Westelinck et al., 2005) |
| P19. Semantic transparency | (De Westelinck et al., 2005; Moody, 2009) |
| P20. Weak coupling can be beneficial | (Zugal, 2013) |
| G25. Model the diagram breadth first, drill down afterwards | (Berenbach, 2004) |
| G26. For low-interactive diagrams, make relations explicit | (Caillies et al., 2002; Rockwell & Bajaj, 2005) |
| G27. For high-interactive diagrams, use text segmentation | (Zugal, 2013) |
| G28. Activate prior knowledge | (Sweller et al., 1998; Van Merriënboer & Kester, 2005) |
| G29. Provide an overview | (Ormrod, 1999; Spence, 2007) |
| G30. Use familiar modelling languages | (Ambler, 2005; De Westelinck et al., 2005; Sweller et al., 1998) |
| G31. Make use of reusable patterns | (Van Merriënboer & Kester, 2005) |
| G32. Use extreme words, visualizations, and examples | (Boyle & Encarnacion, 1998) |
| G33. Use annotations | (Rockwell & Bajaj, 2005; Van Merriënboer et al., 2002; Van Merriënboer & Kester, 2005) |

## 4.8 Perceptual discriminability

The more symbol instances a diagram has, the more difficult it becomes to discriminate between the critical and the non-critical information elements (**Proposition 21**), and the higher the extraneous load (cf. *Proposition 03*) (Paas et al., 2004). When one can omit or hide certain pieces of information (cf. *Guideline 02*), a distinction between core and details is facilitated, which increases diagram effectiveness (Ellis & Dix, 2007; Koshman, 2010). However, in the case where completeness is judged to be key, one may rather put certain information in a complementary text (**Guideline 34**). This strategy of textual differentiation should only be used for high-complexity information though, as for low-complexity information split-attention effects will occur (cf. *Proposition 05*) (Cierniak et al., 2009). Also, decomposable modular structures (**Guideline 35**) allow to first only show the high-level elements (high perceptual discriminability), and afterwards increase the level of refinement (lower perceptual discriminability) (cf. *Section 4.5*). Further, it is argued that *uniform density of nodes* improves diagram quality, i.e., the placement of instances should be spread over the full display area to increase perceptual discriminability (**Guideline 36**) (Schuette & Rotthowe, 1998).

As discussed before, techniques of chunking and spatial proximity can also be used to increase discriminability between different diagram parts (cf. *Section 4.4*) (Paas et al., 2004). Practically, following the laws of the Gestalt Theory (cf. *Proposition 12*), increasing perceptual discriminability of separate chunks can be done by using barrier elements to spatially separate them, such as lines or container elements (**Guideline 37**) (Ding & Mateti, 1990; Koshman, 2010). Distance between diagram elements ensures that the mind perceives them as separate identities (Ding & Mateti, 1990). Therefore, literature in *Information Systems & Computer Science* prompts to avoid spatial *overlap* of shapes (**Guideline 38**) and the crossing of lines (**Guideline 39**) (Ambler, 2005; Purchase, 2002; Ware et al., 2002). Additionally, literature *in Information Systems & Computer Science* argues to maintain a minimum angle between crossing lines (**Guideline 40**), and to fix and spread nodes on an imaginary 'orthogonal unit grid' to *maximize node orthogonality* (cf. *Guideline 23*) (Purchase, 2002; Schuette & Rotthowe, 1998; Ware et al., 2002). Also, diagonal or curved lines lead to lower diagram discriminability and should therefore be avoided (**Guideline 41**) (Ambler, 2005; Ware et al., 2002).

Another guideline is to make sure that the depiction size is reasonable (**Guideline 42**) for proper cognitive perception and reading with moderate effort. Moreover, Ambler (2005) recommends to keep consistent sizing of graphical symbols in order to not draw unwanted attention to larger instances (**Guideline 42**). However, research in *Cognitive & Behavioral Science* has found that diagrams with elements having less differentiable shapes take longer and are harder to process (Peebles & Cheng, 2003). Therefore, a diagram designer should use a visual notation with distinctive symbol shapes (**Guideline 43**) to increase readability of the diagram (Ding & Mateti, 1990). If the notation has low discriminability in symbol shapes, color can be used to make the difference between the object *types* more apparent (**Guideline 44**) (Moody, 2009). In fact, in *Data Science*, colors are preferred to differentiate between item groups in datasets (Koshman, 2010; Ware, 2004). For users who are not familiar with the visual notation, it is suggested to offer a diagram legend that shows all symbols used (visual syntax), together with their meaning (semantics) (**Guideline 45**) (Ambler, 2005).

Table 8. Propositions (P) and guidelines (G) about perceptual discriminability found in literature

| Propositions and Guidelines | Sources |
| --- | --- |
| P21. Core versus detail | (Paas et al., 2004) |
| G02. Filter the represented content | (Mautone & Mayer, 2007; McNamara et al., 1996; Ormrod, 1999; Van Merriënboer & Kester, 2005) |
| G23. Maximize edge and node orthogonality | (Purchase, 2002; Schuette & Rotthowe, 1998) |
| G34. Consider using textual differentiation | (Cierniak et al., 2009) |
| G35. Use modularization | (Ambler, 2005; Gerjets et al., 2004; Regnell et al., 1996) |
| G36. Spread nodes uniformly over the display space | (Schuette & Rotthowe, 1998) |
| G37. Use barriers to separate object groups | (Ambler, 2005; Cox, 1996; Ding & Mateti, 1990) |
| G38. Provide sufficient distance between diagram elements | (Buagajska, 2003; Ding & Mateti, 1990) |
| G39. Avoid close lines, spatial overlap, and line crossings | (Ambler, 2005; Purchase, 2002; Ware et al., 2002) |
| G40. Warrant minimum angle between crossing lines | (Purchase, 2002) |
| G41. Avoid diagonal or bended lines | (Ambler, 2005; Purchase, 2002) |
| G42. Use reasonable and consistent element sizing | (Ambler, 2005; Koshman, 2010) |
| G43. Use distinctive shapes and use them consistently | (Peebles & Cheng, 2003) |
| G44. Use colors to increase element distinctiveness | (Cox, 1996; Ding & Mateti, 1990; Moody, 2009) |
| G45. Use labels and diagram legends | (Ambler, 2005; Moody, 2009) |

## 4.9 Pragmatic clarity

The principle of congruence argues that format and structure of diagrams should correspond to the content they represent (**Proposition 22**) (Tversky et al., 2002). First, the diagram needs to be able to syntactically represent all the semantical constructs of the problem (Cox & Brna, 1995). Therefore, a diagram should use a visual notation that offers clear depictions of its semantical constructs (i.e., be *semantically transparent*) and that also has *semiotic clarity* (i.e., assuring a 1:1 relationship between visual syntax and semantics) (**Guideline 46**) (Berenbach, 2004; Moody, 2009). Next, a diagram designer should first decide on the different semantical constructs that are needed, before picking a visual notation (**Guideline 47**) (Ambler, 2005).

Second, the diagram should have a clear visual structure (**Guideline 48**) to reflect the relationships between diagram elements in a salient, semantically transparent way (Cox & Brna, 1995). Practically, the diagram should have: (1) an obvious and single 'entry point' (**Guideline 49**) (Berenbach, 2004), (2) clear and continuous path flows (**Guideline 50**) (Ding & Mateti, 1990; Field et al., 1993), (3) moderate length of paths (**Guideline 51**) (Ware et al., 2002), (4) minimal length of lines and arrows (**Guideline 51**) (Schuette & Rotthowe, 1998), (5) minimal bending of lines (cf. *Guideline 41*) (Ding & Mateti, 1990), (6) verticality in hierarchal structures (cf. *Guideline 23*) (Schuette & Rotthowe, 1998), (7) maximal edge and node orthogonality (cf. *Guideline 23*) (Ware et al., 2002), and (8) top-bottom ordering for high-to-lower levels (cf. *Guideline 24*) (Ding & Mateti, 1990).

Third, visual syntax should be consistently applied across the diagram (**Guideline 52**) to achieve syntactical clarity (Ambler, 2005). More in particular, literature in *Information Systems & Computer Science* prompts to be consistent in *redundant coding* such as sizing, coloring, fonts, and flow direction (Ambler, 2005; Ding & Mateti, 1990), which can be explained in a diagram legend (cf. *Guideline 45*) (Berenbach, 2004). Moreover, to increase usability and fasten lookup, choose a fixed place on the diagram (like a corner, or at the bottom) to locate the legend and place it there consistently across diagrams (**Guideline 53**) (Ambler, 2005).

Lastly, naming and labelling is an important part of a graphical representation (Irani & Ware, 2003) and can ease navigation between or within diagrams (Berenbach, 2004). When labelling, use concrete and explicit names (**Guidelines 54**). Use verb labels instead of nouns (Amar & Stasko, 2004; Ambler, 2005). Diagram designers should be consistent in the use of domain terminology and naming (Purchase, 2002).

Table 9. Propositions (P) and guidelines (G) about pragmatic clarity found in literature

| Propositions and Guidelines | Sources |
| --- | --- |
| P22. Principle of congruence | (Cox, 1996; Tversky et al., 2002) |
| G23. The diagram should have edge and node verticality | (Ware et al., 2002) |
| G24. The diagram should have top-bottom ordering | (Purchase, 2002) |
| G41. The diagram should have minimal bending of lines | (Ambler, 2005; Purchase, 2002) |
| G45. Provide a diagram legend | (Berenbach, 2004) |
| G46. Choose a semantically transparent notation | (De Westelinck et al., 2005; Lowe, 2003) |
| G46. Choose a semiotically clear notation | (Cox, 1999; De Westelinck et al., 2005; Moody, 2009) |
| G47. First focus on content, then on appearance | (Ambler, 2005) |
| G48. Ensure clarity in structure | (Cox & Brna, 1995) |
| G49. The diagram should have a single and clear entry point | (Berenbach, 2004) |
| G50. The diagram should have symmetry | (Field et al., 1993; Ware et al., 2002) |
| G51. The diagram should have moderate length of path | (Schuette & Rotthowe, 1998; Ware et al., 2002) |
| G52. Be consistent in the use of visual syntax | (Ambler, 2005; Van Merriënboer & Kester, 2005) |
| G53. Be consistent in diagram legend placement | (Ambler, 2005; Berenbach, 2004) |
| G54. Use concrete and explicit naming and labelling | (Amar & Stasko, 2004; Ambler, 2005) |

## 4.10 Visual expressiveness

Whereas the ideal number of elements in a diagram is limited by the working memory (Cowan, 2010; Miller, 1956), the sophistication, or *visual expressiveness*, of the elements is not (**Proposition 23**) (Sweller et al., 1998). By using visual variables, meaning is added without aggravating intrinsic cognitive load (i.e., cognitive processing), as visual variables impact extraneous load (i.e., perceptual processing) (cf. *Proposition 03*).

*Structural object perception theory* states that in the first phase of human *object recognition*, diagrams are scanned for the visual variables that segment the image, to facilitate information extraction (**Proposition 24**) (Irani & Ware, 2003). Bertin (1981) distinguishes eight visual variables, which can be split into *planar variables* and *retinal variables* (**Proposition 25**). The eight variables can be applied on the level of a point, line, or area (cf. *0*) (Roberts, 2000). Various authors argue that designers should use the full scope of visual variables to maximize diagram expressiveness (Ding & Mateti, 1990; Moody, 2009).

The lower the degrees of visual freedom, the less room for interpretation or confusion from the user side (De Westelinck et al., 2005).

The *size* of diagram elements should be balanced, in correspondence to the total diagram size, the number and types of symbols, sizes of other symbols, symmetry and regularity requirements (**Guideline 55**) (Ding & Mateti, 1990). Sizing should be consistent (cf. *Guideline 52*) to maintain readability and to avoid *visual noise effects* (users extracting meaning from unintentionally differing sizes) (Ding & Mateti, 1990).

The *brightness* value reflects the color intensity of an element (Bertin, 1981), which allows for revealing order in information (**Guideline 56**) (Roth & Mattis, 1990). In *Earth Science*, brightness value is used to indicate specific types of clouds to ease satellite image interpretation (Ogao & Kraak, 2002).

The use of *color* is promoted by *Learning & Instruction* literature to discriminate between different exemplar types (**Guideline 57**) (Ormrod, 1999), while *Data Science* frequently uses color to mark patterns, indicate icon attributes, and increase diagram expressiveness (Koshman, 2010). *Computer Science* guidelines prompt to use color to improve readability of diagrams (Ambler, 2005; Coad et al., 1999).

It is recommended to use *shape* to distinguish element types (**Guideline 58**). Because different shapes have different meanings for users with different prior knowledge, the use of more 'standard shapes' like circles, rectangles, and triangles are recommended unless specific (domain-related) shapes are required (Ding & Mateti, 1990). The variable *shape* is one of the most frequently used, and the sole retinal variable in many notations, but it is found to be the least expressive one of all visual variables because it can only represent information on a nominal scale and it is cognitively the least effective (Lohse et al., 1995; Moody, 2009).

*Texture* can be used to eliminate the need for labelling and therefore decrease the cognitive load of a graphical representation (Irani & Ware, 2003). Whereas texture has found to be only marginally effective for interval and ordinal scope of measurement, it is recommended as a good and effective retinal variable for objects measured on a nominal scale (**Guideline 59**) (Pilgrim, 2003).

Experiments in the field of *Vision Research* have shown that it is beneficial to be consistent with *orientation* within and across diagrams to increase readability (**Guideline 60**) (Field et al., 1993). One remark to make here, is to align with the target audience. For example, some users are used to vertically oriented arrays, while others to horizontal arrays (Ding & Mateti, 1990). Additionally, orientation can be used to segment diagram elements into groups (Field et al., 1993). According to the *Gestalt Theory*, elements with the same orientation seem to exert some kind of 'common fate' and are therefore perceived as a group (Koffka, 2013).

Finally, *horizontal* and *vertical positioning* of elements can be used to build hierarchical structures and reflect inter-element relationships (**Guideline 61**). The positioning of elements also determines the symmetry of the diagram (cf. *Guideline 50*), which can improve readability (Schuette & Rotthowe, 1998; Ware et al., 2002).

Following the *principle of congruence* (i.e., "*form follows content*", cf. *Proposition 22*), the semantics should determine the designer's choices in the visual variables that are not specified by the chosen visual language (Ambler, 2005; Ding & Mateti, 1990; Tversky et al., 2002). Moreover, research in *Cognitive Psychology* shows that a user is actually conscious of only a part of the information in working memory, determined by *attention* (**Proposition 26**) (Rowlatt, 2008). Focusing attention has positive effects on information transfer, knowledge compilation, and generating relational statements (Mautone & Mayer, 2007; McAndrew, 1983; Paas et al., 2004; Van Merriënboer & Kester, 2005). *Learning & Instruction* literature argues that intensity of visual variables can be used to guide *attention* of the user (**Guideline 62**) (Ormrod, 1999). Finally, novelty and unexpectedness can create attention effects as well (**Guideline 63**). When an element is initially unexpected within the context, it creates a surprise effect (disfluency effect), increasing attention and retention of that element (**Theory 27**) (Ormrod, 1999).

Table 10. Propositions (P) and guidelines (G) about visual expressiveness found in literature

| Propositions and Guidelines | Sources |
|---|---|
| P23. Principle of sophistication | (Ding & Mateti, 1990; Sweller et al., 1998) |
| P24. Structural object perception theory | (Irani & Ware, 2003) |
| P25. Visual variables of Bertin | (Bertin, 1981) |
| P26. Principle of attention | (Rowlatt, 2008) |
| P27. Disfluency effect | (Ormrod, 1999) |
| G50. Maximize diagram symmetry | (Purchase, 2002; Schuette & Rotthowe, 1998) |
| G52. Be consistent in the use of visual syntax | (Ambler, 2005; Van Merriënboer & Kester, 2005) |
| G55. Balance the size of elements | (Ding & Mateti, 1990) |
| G56. Vary in brightness | (Pilgrim, 2003) |
| G57. Use color to differentiate between element types | (Ormrod, 1999) |
| G58. Use shape to distinguish element types | (Ding & Mateti, 1990) |
| G59. Use texture to differentiate between element types | (Irani & Ware, 2003; Pilgrim, 2003) |
| G60. Be consistent in element orientation | (Field et al., 1993) |
| G61. Attach importance to horizontal and vertical position | (Schuette & Rotthowe, 1998; Ware et al., 2002) |
| G62. Use intensity to signal critical elements | (Ormrod, 1999) |
| G63. Use novelty/unexpectedness to increase expressiveness | (Ormrod, 1999) |

## 4.11 Differentiation

The *Cognitive Fit Theory* (CFT) states that performance is improved when task material representation matches with the task to be executed and with the user performing the task (**Proposition 28**). For example, based on experiments in *Computer Science* it is proposed that novices are better off with graphical representations than with text (**Guideline 64**), while for experts no difference in performance was found (Chan Lin, 2001; Reimann, 2003). Second, accommodating the graphical representation for the learners' prior knowledge level is important to keep motivation high (cf. *Proposition 08*) (Cennamo, 2016; Van Merriënboer & Sweller, 2005). *Learning & Instruction* researchers argue that knowing prior knowledge levels is crucial to determine the right level of element interactivity and refinement to communicate information (Sweller et al., 1998; Van Merriënboer, 1997).

Practically, for novices, it may be a good idea to start with concrete concepts and examples before moving towards a more general, abstract level (**Guideline 65**) (Van Merriënboer et al., 2002). This is called an *inductive-expository strategy* (**Proposition 29a**). For experts however, information can be represented at a higher level of abstraction (i.e., a *deductive strategy*, **Proposition 29b**) (Van Merriënboer et al., 2002). Moreover, the *expertise reversal effect* describes how redundant information (which may be needed for novices) increases cognitive load for experts and decreases performance (**Proposition 30**) (Yeung et al., 1999). Therefore, when the target audience is one of experts, graphical representations should not be overly detailed (**Guideline 66**) (McNamara et al., 1996; Sweller et al., 1998). Additionally, research found that reduced information provision prompts the advanced learner to process more deeply and actively, resulting in positive learning effects (McNamara et al., 1996; Yeung et al., 1999).

Users from different fields have different expectations and conventions regarding lay-out, notation, naming, meta-models etc. (Becker et al., 2000). Therefore, diagram designers may consider the audience familiarity with visual syntax and visual languages (**Guideline 67**). An illustration of this is that programmers generally like to work with rectangular shapes, while researchers in the field of *Graphics* prefer to work with circular shapes (Ding & Mateti, 1990). However, the most conventional or familiar representation mode is not always the best choice. The designer should make a trade-off between the cost of the user's adaptation effort with the potential benefit of higher computational offloading effects (Peebles & Cheng, 2003).

Differences in learning and cognitive abilities of the user should be taken into consideration as well (Cox et al., 1994). Research in *Cognitive Science* has found that high performing students profit from graphically formatted instruction, whereas lower performing students profit from textual instruction (**Guideline 68**) (Cox et al., 1994). Next, the effectiveness of reasoning with diagrams is affected by instructional preferences of the user (Cox, 1996). Depending on where individuals are situated along the visualizer-verbalizer (VV) dimension of cognitive style (**Proposition 31**), they differ in the way they reason with diagrams and are able to distract information from different formats (Cox & Brna, 1995). A diagram designer should differentiate his or her design for the cognitive style of the target audience (**Guideline 69**).

Further, learner performance is improved when the structure and representation format of the learning material corresponds with the task requirements (cf. *Proposition 28*) (Vessey & Galletta, 1991). Depending on the diagramming purpose, different graphical design principles should thus be prioritized (**Guideline 70**). For example, whereas diagrams built for specification purposes should aim for exhaustive and precise visualization modes, diagrams for communication should rather focus on clarity and understandability.

Finally, diagram designers should evaluate visualization techniques for their ability to represent the semantics of the domain and to facilitate understanding of the domain (Gemino & Wand, 2003). Depending on the specific semantical requirements of the information to be presented, the designer should differentiate the visualization strategy (**Guideline 71**). Studies in the field of *Learning & Instruction* found that depictive formats (e.g., flowcharts) are better suited to represent abstract and complex information, whereas descriptive formats (e.g., verbal or textual) are better for more concrete, lower-complexity information (Cox, 1999; Schnotz, 2013).

Table 11. Propositions (P) and guidelines (G) about differentiation found in literature

| Propositions and Guidelines | Sources |
|---|---|
| P28. Cognitive Fit Theory | (Ainsworth, 2006; Gilmore & Green, 1984; Vessey & Galletta, 1991) |
| P29. Inductive-expository strategy and deductive strategy | (Van Merriënboer et al., 2002) |
| P30. Expertise reversal effect | (Sweller et al., 1998) |
| P31. Visualizer-verbalizer (VV) dimension of cognitive style | (Cox, 1996; Cox & Brna, 1995) |
| G64. Use graphical representations for novice audiences | (Chan Lin, 2001; Reimann, 2003) |

| Propositions and Guidelines | Sources |
|---|---|
| G65. Use gradual content build-up for novice audiences | (Van Merriënboer et al., 2002) |
| G66. Do not represent too much details for expert audiences | (McNamara et al., 1996; Yeung et al., 1999) |
| G67. Consider familiarity with visual languages | (Becker et al., 2000; Peebles & Cheng, 2003) |
| G68. Prefer graphical formats for high-performing students | (Cox et al., 1994) |
| G68. Prefer textual formats for lower performing students | (Cox et al., 1994) |
| G69. Differentiate for cognitive styles of users | (Claes et al., 2017; Cox, 1996) |
| G70. Differentiate for task type | (Vessey & Galletta, 1991) |
| G71. Differentiate for semantical requirements | (Gemino & Wand, 2003) |

## 4.12 Balancing

Graphical representation design is not about insisting on extremes (Purchase, 2002), but rather about making purposeful trade-offs (**Guideline 72**) between various low-level design objectives (e.g., *clarity*, *expressiveness*, and *completeness*) in order to achieve the high-level diagram objective (e.g., effective information transfer to learners with low prior knowledge). This challenging balancing exercise has a high impact on the diagram quality, yet little literature exists on this topic. The lack of clear balancing conventions and uniformity in design processes causes designers to build diagrams according to their preferences or habit, often leading to suboptimal and ineffective graphical representations (Berenbach, 2004).

Purchase proposes seven metrics to quantify the extent of aesthetic conformance of a graphical representation (Purchase, 2002). For each rule, a formal algorithm calculates a score between 0 (zero conformance) and 1 (full conformance). This way, the aesthetic conformance of diagrams is measured from different perspectives and on a continuous spectrum (**Guideline 73**). Depending on the target audience and the intended tasks, weights can be assigned to the different guidelines to indicate and formalize relative importance (Ding & Mateti, 1990). For example, *Guideline 39* ('minimize crossings') can be given a higher priority than *Guideline 51* ('minimize total path length'). It is important to remark here that the weighting of rules will then add a subjective notion to this so far objective evaluation methodology.

The *multimedia theory* (**Proposition 32**) proposes to offer multiple representation formats (**Guideline 74**), which allows the user to choose the best one for the task at hand (Ainsworth, 2006; Dunn & Dunn, 1993). Additionally, a *single* diagram can still be 'multimedia' through combining depictive formatting with text (**Guideline 75**). Mayer's *multimedia effect* states that learners learn more from the combination of words and graphics together than from words alone (Moreno & Mayer, 1999; Van Merriënboer & Kester, 2005). However, research on the effectiveness of the multimedia effect has resulted in contradicting results: whereas various studies found positive results (Cox & Brna, 1995; Mayer & Sims, 1994), several others did not (Ainsworth, 2006; Chandler & Sweller, 1992). For example, studies on analogical reasoning show that learners' understanding improves from comparing different information sources (Gentner & Markman, 1997). But various other studies have observed that users have difficulties with integrating information from multiple sources and tend to treat multiple representations in isolation (Ainsworth, 2006) (cf. *split-attention effect*, *Proposition 05*).

In his *integrated model of text and image comprehension* (**Proposition 33**), Schnotz (2013) argues that there are conditional principles for the use of multimedia. He argues that (1) combining diagrams and text is only effective when learners have low prior knowledge but enough cognitive abilities to process both the depictive and descriptive information (**Guideline 76**), (2) depictive and descriptive formats should not be combined if one or the other is redundant (**Guideline 77**), (3) pictures should only be combined with text when their semantical relation is clear (**Guideline 78**), and (4) text and graphical representations should be presented in close spatial proximity (**Guideline 79**). Multimedia can thus be used to improve learning from diagrams, but only under the stated conditions.

Although there are many visual variables and techniques that were found to improve diagram expressiveness (Lorch et al., 1993; Reynolds & Shirey, 1988), finding the right balance between all of them is challenging. Too much focusing will lead to *tunnel vision* effects, yet too little focus will lead to *splatter vision* (**Proposition 34**) (Fiol & Huff, 1992). While multiple authors recommend using the full range of visual variables, *Learning & Instruction* guidelines warn not to overload the diagram with retinal variables (**Guideline 80**). Novices get easily distracted by prominent symbols and may miss the semantics behind the syntax (Lowe, 2003). Also, redundant symbols unnecessarily increase cognitive load and may decrease diagram readability (Schnotz, 2013). Some research even argues that instead of trying to maximize visual expressiveness, diagram design should minimize the expressiveness by maximizing construct specificity and decreasing the need for retinal variables (**Guideline 81**) (Reimann, 2003). We recommend making the trade-off between visual expressiveness and diagram readability depending on the particular diagramming goal.

Table 12. Propositions (P) and guidelines (G) about balancing found in literature

| Propositions and Guidelines | Sources |
|---|---|
| P32. Multimedia theory and multimedia effect | (Schnotz, 2013; Van Merriënboer & Kester, 2005) (Ainsworth, 2006; Moreno & Mayer, 1999) |
| P33. Integrated model of text and image comprehension | (Schnotz, 2013) |
| P34. Splatter vision versus tunnel vision effect | (Fiol & Huff, 1992) |
| G72. Balance the guidelines to multiple users and tasks | (Purchase, 2002) |
| G73. Measure diagram effectiveness by continuous metrics | (Purchase, 2002) |
| G74. Use multimedia to support multiple users and tasks | (Ainsworth, 2006; Dunn & Dunn, 1993) |
| G75. Combine depictive (graphics) and descriptive (text) | (Moreno & Mayer, 1999; Van Merriënboer & Kester, 2005) |
| G76. Only combine graphics and text for low expertise users | (Schnotz, 2013) |
| G77. Only combine graphics and text if neither is redundant | (Schnotz, 2013) |
| G78. Only combine graphics and text if their relation is clear | (Schnotz, 2013) |
| G79. Represent graphics and text in close spatial proximity | (Schnotz, 2013) |
| G80. Use the principle of expressiveness with moderation | (Reimann, 2003) |
| G81. Maximize construct specificity | (Reimann, 2003) |

# 5 Design principles for graphical representation design

The literature study resulted in the identification of 34 propositions and 81 guidelines about graphical representation design, grouped in 12 focus topics. Based on the logical coherence of the 12 focus topics, the underlying propositions and guidelines were aggregated into seven foundational principles, divided into three categories. Fig. 4 shows an overview of these principles and categories, together with the focus topics to which they relate. They are presented and discussed in this section and they jointly form the "*Physics of Diagrams*" proposed in this paper.

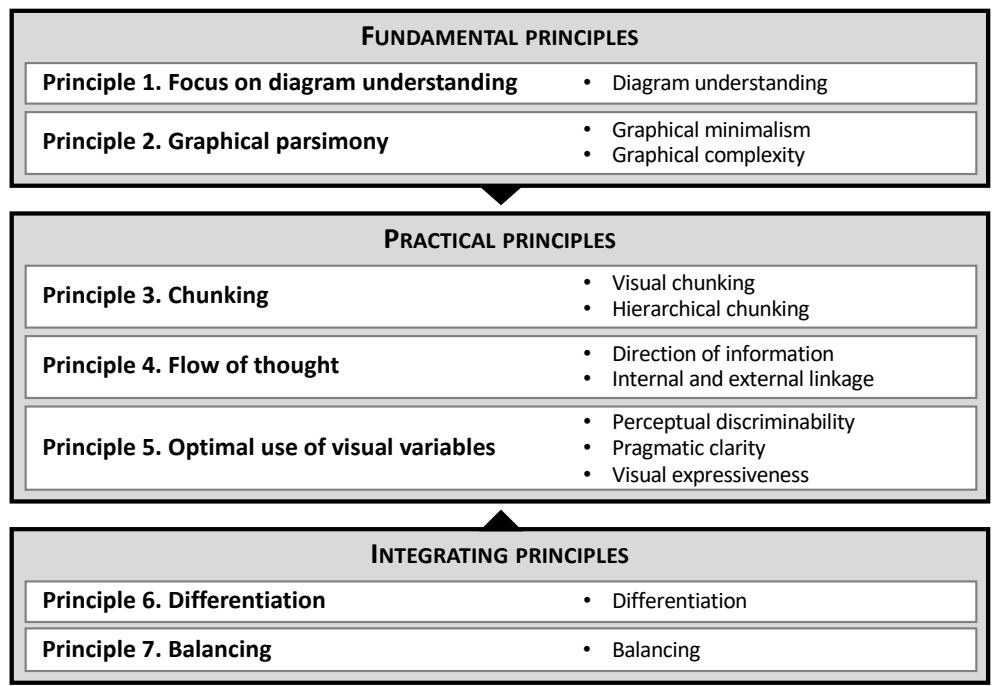

Fig. 4. The "*Physics of Diagrams*": Seven fundamental principles for graphical representation design

## 5.1 Fundamental principles

The *fundamental principles* provide the foundation for and the reasoning behind the practical and integrating principles. Two fundamental principles were identified: the principle of focus on diagram understanding and the principle of graphical parsimony.

**Principle 1.   Focus on diagram understanding**

> The **principle of focus on diagram understanding** states that a diagram designer should focus on diagram understanding and the ease of cognitive processing of the diagram.

In the past, quality frameworks for graphical conceptual models have been characterized by a vigorous pursuit for completeness and correctness. Two illustrations of this are the leading quality frameworks of Lindland, Sindre, and Sølvberg (LSS) (Lindland et al., 1994) and Bunge, Wand, and Weber (BWW) (Wand & Weber, 1990). However, immoderate focus on completeness and correctness often translates into complex diagrams, which increases cognitive load (cf. *Proposition 03*) and trigger split-attention effects (cf. *Proposition 05*). These diagrams have low readability and thus fail to achieve their purpose: facilitating and enhancing conveyance of information (Cierniak et al., 2009; Florax & Ploetzner, 2010; Ginns, 2006; Rockwell & Bajaj, 2005; Sweller et al., 1998; Zugal, 2013). When this happens, users also benefit less from the efforts to optimize completeness and correctness. Consequently, in order for graphical representations to fulfil their raison d'être, diagram designers should also focus on assuring diagram understanding, balancing concerns of ease of cognitive processing and the completeness and correctness of the diagram.

**Principle 2.   Graphical parsimony**

> According to the principle of graphical parsimony, graphical representations should contain the minimum possible amount of symbol types and of symbol instances of each type.

Due to the growing complexity of data (i.e., an increasing level of abstraction and interconnectedness of data), which is aggravated by the growing volume, variety, velocity, and veracity of data (i.e., 'big data'), visual representations often get overcrowded (Ellis & Dix, 2007). Too much information needs to be fit on a physically and cognitively constrained display space. The issue of *diagram overcrowding* is two-sided: on the one hand, it is caused by an excessive amount of *types* of graphical symbols on the diagram, on the other hand, by an exaggerated number of symbol *instances*. Visualization techniques need to adopt strategies for dealing with both sides of overcrowding. The first way is to adhere to the notion of *graphical minimalism* (minimal number of symbol *types*) (Si-said Cherfi et al., 2002), the second and complementary way is to adhere to the notion of *graphical complexity management* (minimal number of symbol *instances*).

**Illustration of the principle of graphical parsimony in practice.** In *Information Systems & Computer Science*, BPMN model designers often construct diagrams according to the 7 Process Modelling Guidelines (7PMG) (Mendling et al., 2010) and the guidelines of Bruce Silver (Silver, 2011). The resulting diagrams contain as few element instances as possible, have only one start and end event, and fit onto one page (models with more than 50 elements are decomposed). One of the most illustrative examples can be found in the *Transport* sector where typical public transport maps are stripped from all unnecessary details, presenting various available transport routes as simplified, stylized colored lines.

## 5.2   Practical principles

The *practical principles* build on the fundamental principles and describe more concrete how to achieve diagram effectiveness. Three practical principles were identified: the principle of chunking, the principle of flow of thought, and the principle of optimal use of visual variables.

**Principle 3.   Chunking**

> The **principle of chunking** states that information on a graphical representation should be grouped into meaningful parts to improve understanding and diagram acquisition. The cognitive load of diagrams is further decreased by using hierarchical structures to represent element relationships.

Some benefits of chunking are named in literature. First, clear visual-spatial characteristics (i.e., structural coherence) have been found to help see relationships between elements (Hall & O'Donnell, 1996; Krajcik, 1991; Linn & Muilenburg, 1996) and support learning from graphical representations (De Westelinck et al., 2005; Lowe, 2003; Paas et al., 2004; Saltz, 1971). This is especially the case for information of high cognitive load (Cox, 1999; Haygood & Bourne, 1965; Mayer, 1976; Neisser & Weene, 1962) and when the user is a novice (Sweller, 1994). Second, research in *Cognitive Psychology* has found that creative thinking with diagrams requires a hierarchical structure and an easy identification of cross-links (Novak & Cañas, 2008), which chunking can provide. Next, research in *Educational Psychology* found that users who received structural graphical representations were able to formulate more relational (however not necessarily more causal) statements than the users in the control group (Mautone & Mayer, 2007). Furthermore, hierarchical diagrams allow for *simultaneous learning*. In the case of high-element interactivity, simultaneous display is often required to support simultaneous processing and manipulation (Bourne et al., 1971; Ginns, 2006; Paas et al., 2004). Lastly, chunking positively impacts the maintainability of the diagram. For example, in *Information Systems & Computer Science*, software is modularized in packages to improve maintainability (Bavota et al., 2010). In short, the positive impact of chunking appears to be very stable over a range of different conditions such as different measures of learning time, low versus high prior knowledge, varying learning tasks and the extent of additional explanations (Gerjets et al., 2004).

**Illustration of the principle of chunking in practice.** In *Information Systems & Computer Science*, designers often use container elements, i.e., a combination of spatial proximity and visual variables, to spatially group information together. An illustration hereof is the use of lanes in business process diagrams to distinguish different roles involved in process execution. In *Software Engineering*, modularization is used to facilitate and improve software design, coding, and software maintainability. In order to be able to modularize, engineers need a visual notation that offers the necessary semantical constructs to do so (Ågerfalk et al., 2007; Moody, 2009). Examples of *modularization constructs* are 'sub-system constructs' (like sub-processes seen in business process diagrams) (De Meyer & Claes, 2018; Sadowska, 2013) and 'decomposable constructs' (like packages in Object Oriented software, which group related classes together) (Bavota et al., 2010). Although literature in various fields agrees on the desirability of modularization (Bavota et al., 2010; De Meyer & Claes, 2018; Mautone & Mayer, 2007; Saltz, 1971), many leading visual notations still lack the required modularization constructs. An example is i*, a leading visual notation for goal modeling in requirements engineering, which does not allow for proper modularization as it does not offer the required semantical constructs.

**Principle 4.    Flow of thought**

> The **principle of flow of thought** states that the readability of a graphical representation is improved by maintaining a continuous and consistent information flow direction and through internal and external linking of the diagram elements.

In *Cognitive Sciences*, research about attention revealed that humans think in an endless stream of connected concepts, which is often referred to as the "*stream of consciousness*" or the "*flow of thought*" (James, 1983). However, when a reader "*loses the train of thought*", comprehension is lost (Morris, 1901) and the primary goal of diagram understanding is hindered. Therefore, it is important to help a diagram reader through the complexity of the represented material by keeping a logical, consistent information flow direction (Field et al., 1993), as well as by providing visual links between related concepts within the diagram (internal linkage) and to the relevant external environment of the diagram (external linkage) (Houston, 1965; Saltz, 1971; Sweller, 1988). External linkage can for instance be achieved by activating prior knowledge and by making use of the principle of *repleteness* (cf. *Proposition 18*) (Berenbach, 2004).

**Illustration of the principle of flow of thought in practice.** An example in the field of *System Engineering* is the N-squared chart, where input data enters on the top-left side of the diagram, whereas output leaves at the bottom-right corner (Long, 2002). In the field of *Project Management*, baseline schedules take on a left-to-right, top-to-bottom approach (Vanhoucke, 2012). In *Mathematics*, a chart containing multiple data series is often represented as a line chart, such that the different data points within a single series are internally linked. External linking often refers to the concept of semantic transparency, which is a technique to use symbols of which the meaning can be guessed by novice users through the mapping to familiar concepts. For example, in the *Energy* sector, in power grid diagrams, the color red is often used where capacity is (at risk of being) insufficient (e.g., in (Smith, 2002)).

**Principle 5.   Optimal use of visual variables**

> The **principle of optimal use of the visual variables** states that graphical representation readability can be improved by making diagram elements more perceptually discriminable; by ensuring clarity in semantics, syntax, naming, and diagram structure; and through the expressive use of visual variables.

Human processing of a graphical representation takes place in two phases: perceptual processing (seeing) and cognitive processing (understanding) (Newell & Simon, 1972). Whereas cognitive processing is slow, sequential and under conscious control of attention, perceptual processing is fast, parallel and rather unconscious, because it happens through dedicated feature detectors for color, shape, etc. (Atkinson & Shiffrin, 1968). Therefore, perceptual discriminability plays an important role in the effectiveness and efficiency of understanding a diagram. Further, pragmatic clarity in graphical representation design leads to improved learning by increasing cognitive effectiveness of the diagrams, keeping the user's motivation up, and reducing fear of failure (Cennamo, 2016; Moreno, 2006). Research in *Educational Technology* proposes to achieve clarity in graphical representation design by using simple and clear visual notation (Cennamo, 2016): The meaning of every object type, relational structure and visual variable used on the diagram should be salient. Experiments in the field of *Learning & Instruction* have found that learners often struggle with understanding graphical representations because they have no knowledge of the visual notation or of the general syntax that is used (Cox, 1999; De Westelinck et al., 2005). Moreover, it was found that complex reasoning with unfamiliar visual syntax is prone to be ineffective (Sweller et al., 1998) and that users learn best from diagrams with clear visual properties (Lowe, 2003).

**Illustration of the principle of optimal use of visual variables in practice.** Examples of this principle are abundant in practice. In *Software Engineering*, programming editors use automatic syntax highlighting and lay-outing to support the developers with the complexity of their software code. Next, in *Traffic*, road signs differ in color, shape, size and placement to maximize prompt understanding. Also the design of the dashboard and GPS screen in cars has been optimized using this principle, such that important regular information is clearly visible (e.g., relatively big indication of current speed) and such that issue-related information pops out (e.g., orange, blinking warning signs).

## 5.3 Integrating principles

The last category of principles, the *integrating principles*, indicate how to resolve conflicts between the previously mentioned principles and how to facilitate the evaluation of design choices for different tasks and audiences. The two integrating principles are the principle of differentiation and the principle of balancing.

**Principle 6.   Differentiation**

> The **principle of differentiation** states that diagram readability can be improved by differentiating diagram design for user characteristics, task characteristics, and semantical requirements of the graphical representation.

Different visual-spatial characteristics are appropriate in different contexts (Cox, 1999). The three context factors that influence graphical representation appropriateness (and thus effectiveness) are *user characteristics* (a combination of prior knowledge, cognitive abilities, and cognitive style), *task characteristics* (fit for purpose), and *semantical requirements* (the ability of the syntax to convey the semantics) (Ainsworth, 2006; Cox, 1999). When these factors are known, graphical representation design can be optimized correspondingly, prioritizing some principles and strategies over others (Ainsworth, 2006).

**Illustration of differentiation techniques in practice.** A new emerging field that is particularly strong in general format differentiation is *Adaptive Learning & Instruction*. Adaptive learning software takes learner's ongoing level of expertise and prior knowledge into account and dynamically adjusts content provision and content format accordingly (Van Merriënboer & Sweller, 2005). This way, graphical representations are optimally differentiated for the user and attain higher levels of effectiveness. In the field of *Business Process Modeling*, it is argued that not only should the diagram itself be differentiated towards the user(s) and the task(s) it has to support (cf. *Proposition 28*), also the creation process of the diagram is more effective and efficient if the approach is fitted to the model designer(s) (Claes et al., 2015).

**Principle 7.   Balancing**

> The **principle of balancing** argues that when a graphical representation must accommodate multiple users and tasks, graphical representation design should use multi-media techniques and assign weights of relative importance to the different former principles of graphical representation design.

In some cases, graphical representations cannot be customized for target audience or tasks (cf. *differentiation*), because they are used by a diverse range of users and for various tasks. Therefore, it is increasingly relevant and important to evaluate the quality of graphical representations from different viewpoints (Becker et al., 2000). These different viewpoints are embodied within the graphical representation design guidelines that are discussed above. Each principle represents a different component and perspective on diagram effectiveness and should be given consideration. While it is never beneficial to completely neglect a graphical design principle, appropriate design rationale can allocate higher priority to certain components without completely disregarding the others (Ding & Mateti, 1990).

# 6 Discussion and conclusion

In an attempt to create a scientific basis for graphical representation design, seven principles for high-quality graphical representation design have been defined, divided into three groups: fundamental, practical, and integrating principles. The fundamental principles provide the foundation for and reasoning behind the practical principles. Theories from *Cognitive Science* emphasize the need for diagram readability (*Principle of focus on diagram understanding*), which leads to the need for graphical minimalism and for complexity management (*Principle of graphical parsimony*). Further, the practical principles indicate *how* diagram design can conform to the fundamental principles and thus achieve diagram effectiveness. Essentially, it is proposed to consider visual chunking and hierarchical modularization (*Principle of chunking*); information direction and linkage (*Principle of flow of thought*); and graphical distinction, pragmatic clarity, and visual expressiveness (*Principle of optimal use of the visual variables*). Finally, the integrating principles indicate how to resolve conflicts between the principles that originate from the fact that diagrams may be used by multiple end users and for multiple tasks (*Principle of differentiation*), and those conflicts that originate from opposing effects of the various principles (*Principle of balancing*).

There are three typical tasks that the proposed principles support. First, having explicit design principles helps to create uniformity in the design process and to formalize and guide design decisions ('*design*'). Second, different diagram designs can be compared by analyzing how well each of them performs against certain design principles ('*comparison*'). Third, the principles can be used for formalizing *quality assurance* of diagram design ('*evaluation*'). Through the use of the quality evaluation technique of Purchase discussed in Section 4.12, the impact of design decisions can be 'quantified'. The metric score system allows a diagram designer to see what the relative impact is of a particular design decision on the conformance to the different principles in a non-binary (yes/no) way.

Various stakeholders can benefit from the application of the principles. Graphical representations are used to communicate to both novice and expert users, and between them. Novices (e.g., customers, top-level management) will be able to better understand the represented information, leading to higher efficiency and less errors. Experts will be able to better collaborate through standardized design processes and formats (Scaife & Rogers, 1996), and profit from diagrams that are easier to adapt to changes in the domain (Nelson et al., 2012; Zugal, 2013). Finally, the consequences of novice-expert differences (e.g., communication problems, expertise reversal effects, reading difficulties for novices) are minimized.

The principles defined in this paper are based upon perceptual and cognitive processing theories, which apply across various domains. Examples throughout the paper already signaled application areas such as *Earth & Environmental Sciences* (e.g., Data Discovery Software), *Project Management* (e.g., Baseline Schedules), *Data Science* (e.g., visual filtering), *Medical Statistics* (e.g., advanced medical data analysis), *Requirements Engineering* (e.g., Goal Models), *Tacit Knowledge Documentation* (e.g., Cognitive Maps), *System Engineering* (e.g., N-squared Charts), *Adaptive e-Learning* (e.g., Adaptive Visualization Techniques), *Cartography* (e.g., geographical maps), and *Transportation Industry* (e.g., Risk Analysis Models).

The research described in this paper comes with a number of limitations. This paper does not dictate what type of graphical representation (graphs, tables, decision trees, flow charts, etc.) one should use for defined types of scenarios. It only hints towards what type of visualization strategy matches best with the type of information or with the target audience. No exhaustive overview or evaluation of the existing graphical representation techniques and strategies is provided. Further, this paper takes a user-perspective on graphical representation design and focusses merely on these encoding decisions that directly impact the decoding side. Yet, the encoding side, i.e., the designer-perspective, is also important to consider. Ease of the designing and maintainability of the representations are also measures for quality in graphical representation design and are valuable to take into account (Evitts, 2000; Viyović et al., 2014). Lastly, while the literature study is assumed to broadly cover the problem space of the use of visual syntax on a sentence level, it is likely that not all knowledge that is potentially relevant for graphical representation design is covered in this paper. We encourage future research to build further on a scientific basis for graphical representation design and take over where this paper left off.

# References

Ågerfalk, P. J., Brinkkemper, S., Gonzalez-Perez, C., Henderson-Sellers, B., Karlsson, F., Kelly, S., & Ralyté, J. (2007). Modularization constructs in method engineering: Towards common ground? In J. Ralyté, S. Brinkkemper, & B. Henderson-Sellers (Eds.), *Situational Method Engineering: Fundamentals and Experiences, IFIP WG 8.1 Working Conference, Proceedings* (Vol. IFIPAICT 2, pp. 359–368). Geneva, Switzerland: Springer. https://doi.org/10.1007/978-0-387-73947-2_27

Ainsworth, S. (2006). DeFT: A conceptual framework for considering learning with multiple representations. *Learning and Instruction*, *16*(3), 183–198. https://doi.org/10.1016/j.learninstruc.2006.03.001

Amar, R., & Stasko, J. (2004). A knowledge task-based framework for design and evaluation of information visualizations. In *10th IEEE Symposium on Information Visualization, INFOVIS 2004, Proceedings* (pp. 143–149). Austin, TX, USA: IEEE. https://doi.org/10.1109/INFVIS.2004.10

Ambler, S. W. (2005). *The elements of UML™ 2.0 style*. Cambridge University Press. https://doi.org/10.1017/CBO9780511817533

Atkinson, R. C., & Shiffrin, R. M. (1968). Human memory: A proposed system and its control processes. *Psychology of Learning and Motivation*, *2*, 89–195. https://doi.org/10.1016/S0079-7421(08)60422-3

Austin, K. A. (2009). Multimedia learning: Cognitive individual differences and display design techniques predict transfer learning with multimedia learning modules. *Computers and Education*, *53*(4), 1339–1354. https://doi.org/10.1016/j.compedu.2009.06.017

Bavota, G., De Lucia, A., Marcus, A., & Oliveto, R. (2010). Software re-modularization based on structural and semantic metrics. In *17th Working Conference on Reverse Engineering, WCRE 2010, Proceedings* (pp. 195–204). IEEE. https://doi.org/10.1109/WCRE.2010.29

Becker, J., Rosemann, M., & Von Uthmann, C. (2000). Guidelines of business process modeling. In W. M. P. Van der Aalst, J. Desel, & A. Oberweis (Eds.), *Business Process Management. Models, Techniques, and Empirical Studies. Part I.* (Vol. LNCS 1806, pp. 30–49). Springer. https://doi.org/10.1007/3-540-45594-9_3

Berenbach, B. (2004). The evaluation of large, complex UML analysis and design models. In *26th International Conference on Software Engineering, ICSE 2004, Proceedings* (pp. 232–241). Edinburgh, UK: IEEE. https://doi.org/10.1109/ICSE.2004.1317445

Bertin, J. (1981). Graphics and graphic information processing. In *Graphics and Graphic Information Processing* (pp. 24–31). https://doi.org/10.1515/9783110854688

Billington, J., Christensen, S., & Van Hee, K. M. (2003). The Petri Net Markup language: Concepts, technology, and tools. In W. M. P. Van der Aalst & E. Best (Eds.), *24th International Conference on Application and Theory of Petri Nets, ICATPN 2003, Proceedings* (Vol. LNCS 2679, pp. 483–505). Eindhoven, the Netherlands. https://doi.org/10.1007/3-540-44919-1_31

Bourne, L. E., Ekstrand, D. R., & Dominowski, R. L. (1971). *The psychology of thinking*. Oxford, UK: Prentice-Hall.

Bower, G. H. (1970). Organizational factors in memory. *Cognitive Psychology*, *1*(1), 18–46. https://doi.org/10.1016/0010-0285(70)90003-4

Boyle, C., & Encarnacion, A. O. (1998). MetaDoc: an adaptive hypertext reading system. In P. Brusilovsky, A. Kobsa, & J. Vassileva (Eds.), *Adaptive Hypertext and Hypermedia* (pp. 71–89). Springer.

Brophy, J., & Good, T. (1986). Teacher-effects results. In M. C. Wittrock (Ed.), *Handbook of Research on Teaching* (p. 1037). New York: Macmillan.

Buagajska, M. (2003). Classification model for visual spatial design guidelines in the digital domain. In *2nd Workshop on Software and Usability Cross, IFIP WG 13.2, Proceedings* (p. 11).

Caillies, S., Denhière, G., & Kintsch, W. (2002). The effect of prior knowledge on understanding from text: Evidence from primed recognition. *European Journal of Cognitive Psychology*, *14*(2), 267–286. https://doi.org/10.1080/09541440143000069

Carlson, R. A., Khoo, B. H., & Elliott II, R. G. (1990). Component practice and exposure to a problem-solving context. *Human Factors: The Journal of the Human Factors and Ergonomics Society*, *32*(3), 267–286. https://doi.org/10.1177/001872089003200302

Casner, S. (1989). *A task-analytic approach to the automated design of information graphics. Technical Report AlP - 82*.

Cennamo, K. S. (2016). Learning from video: Factors influencing learners' preconceptions and invested mental effort. *Educational Technology Research and Development*, *41*(3), 33–45. https://doi.org/10.1007/BF02297356

Chan Lin, L. J. (2001). Formats and prior knowledge on learning in a computer-based lesson. *Journal of Computer Assisted Learning*, *17*(4), 409–419. https://doi.org/10.1046/j.0266-4909.2001.00197.x

Chandler, P., & Sweller, J. (1992). The split-attention effect as a factor in the design of instruction. *British Journal of Educational Psychology*, *62*(2), 233–246. https://doi.org/10.1111/j.2044-8279.1992.tb01017.x

Chen, C. (2005). Top 10 unsolved information visualization problems. *IEEE Computer Graphics and Applications*, *25*(4), 12–16. https://doi.org/10.1109/MCG.2005.91

Cierniak, G., Scheiter, K., & Gerjets, P. (2009). Explaining the split-attention effect: Is the reduction of extraneous cognitive load accompanied by an increase in germane cognitive load? *Computers in Human Behavior*, *25*(2), 315–324. https://doi.org/10.1016/j.chb.2008.12.020

Claes, J., Vanderfeesten, I., Gailly, F., Grefen, P., & Poels, G. (2015). The Structured Process Modeling Theory (SPMT) - A cognitive view on why and how modelers benefit from structuring the process of process modeling. *Information Systems Frontiers*, *17*(6), 1401–1425.

Claes, J., Vanderfeesten, I., Gailly, F., Grefen, P., & Poels, G. (2017). The Structured Process Modeling Method (SPMM) - What is the best way for me to construct a process model? *Decision Support Systems*, *100*(6), 57–76. https://doi.org/10.1016/j.dss.2017.02.004

Coad, P., Luca, J. D., & Lefebvre, E. (1999). *Java modeling color with UML: Enterprise components and process with Cdrom* (1th ed.). Upper Saddle River, New Jersey: Prentice Hall.

Cowan, N. (2010). The magical mystery four: How is working memory capacity limited, and why? *Current Directions in Psychological Science*, *19*(1), 51–57.

Cox, R. (1996). *Analytical reasoning with multiple external representations*. The University of Edinburgh.

Cox, R. (1999). Representation construction, externalised, cognition and individual differences. *Learning and Instruction*, *9*(4), 343–363. https://doi.org/10.1016/S0959-4752(98)00051-6

Cox, R., & Brna, P. (1995). Supporting the use of external representations in problem solving: The need for flexible learning environments. *Journal of Artificial Intelligence in Education*, *6*(November), 239–302.

Cox, R., Stenning, K., & Oberlander, J. (1994). Graphical effects in learning logic: Reasoning, representation and individual differences. In A. Ram & K. Eiselt (Eds.), *16th Annual Conference of the Cognitive Science Society, Proceedings* (pp. 237–242). Lawrence Erlbaum Associates.

Crosby, P. B. (1979). *Quality is free: The art of making quality certain*. New York: McGraw-Hill.

De Meyer, P., & Claes, J. (2018). An overview of process model quality literature - The Comprehensive Process Model Quality Framework. *ArXiv:1808.07930*, 31 pages.

De Westelinck, K., Valcke, M., De Craene, B., & Kirschner, P. A. (2005). Multimedia learning in social sciences: Limitations of external graphical representations. *Computers in Human Behavior*, *21*(4), 555–573. https://doi.org/10.1016/j.chb.2004.10.030

DeMarco, T. (2002). Structured analysis: beginnings of a new discipline. In M. Broy & E. Denert (Eds.), *Software Pioneers* (pp. 520–527). Springer. https://doi.org/10.1007/978-3-642-59412-0_32

Ding, C., & Mateti, P. (1990). A framework for the automated drawing of data structure diagrams. *IEEE Transactions on*

*Software Engineering*. https://doi.org/10.1109/32.52777

Dunn, R. S., & Dunn, K. J. (1993). *Teaching secondary students through their individual learning styles: Practical approaches for grades 7-12*. Prentice Hall.

Ellis, G., & Dix, A. (2007). A taxonomy of clutter reduction for information visualisation. *IEEE Transactions on Visualization and Computer Graphics*, *13*(6), 1216–1223. https://doi.org/10.1109/TVCG.2007.70535

Evitts, P. (2000). *A UML pattern language* (First Edit). Indianapolis: New Riders Publishing.

Field, D. J., Hayes, A., & Hess, R. F. (1993). Contour integration by the human visual system: Evidence for a local "association field." *Vision Research*, *33*(2), 173–193. https://doi.org/10.1016/0042-6989(93)90156-Q

Figueres-Esteban, M., Hughes, P., & Van Gulijk, C. (2015). The role of data visualization in railway big data risk analysis. In *25th European Safety and Reliability Conference, ESREL 2015, Proceedings* (pp. 2877–2882). CRC Press / Balkema.

Fiol, C. M., & Huff, A. S. (1992). Maps for managers: Where are we? Where do we go from here? *Journal of Management Studies*, *29*(3), 267–285. https://doi.org/10.1111/j.1467-6486.1992.tb00665.x

Fiske, J. (1990). *Introduction to communication studies* (2nd ed.). Routledge. https://doi.org/kommunikation; kommunikationswissenschaft; semiotik; zeichen; mythos; strukturalismus

Florax, M., & Ploetzner, R. (2010). What contributes to the split-attention effect? The role of text segmentation, picture labelling, and spatial proximity. *Learning and Instruction*, *20*(3), 216–224. https://doi.org/10.1016/j.learninstruc.2009.02.021

Gaissmaier, W., Wegwarth, O., Skopec, D., Müller, A.-S., Broschinski, S., & Politi, M. C. (2012). Numbers can be worth a thousand pictures: Individual differences in understanding graphical and numerical representations of health-related information. *Health Psychology*, *31*(3), 286–296. https://doi.org/10.1037/a0024850

Garvin, D. A. (1984). Product quality: An important strategic weapon. *Business Horizons*, *27*(3), 40–43. https://doi.org/10.1016/0007-6813(84)90024-7

Gemino, A. C., & Wand, Y. (2003). Evaluating modeling techniques based on models of learning. *Communications of the ACM*, *46*(10), 79–84.

Gentner, D., & Markman, A. B. (1997). Structure mapping in analogy and similarity. *American Psychologist*, *52*(1), 45–56. https://doi.org/10.1037//0003-066X.52.1.45

Gerjets, P., Scheiter, K., & Catrambone, R. (2004). Designing instructional examples to reduce intrinsic load: Molar versus modular presentation of solution procedures. *Instructional Science*, *32*(1–2), 33–58. https://doi.org/10.1023/B:TRUC.0000021809.10236.71

Ghylin, K. M., Green, B. D., Drury, C. G., Chen, J., Schultz, J. L., Uggirala, A., … Lawson, T. A. (2007). Clarifying the dimensions of four concepts of quality. *Journal of Theoretical Issues in Ergonomics Science*, *9*(1), 73–94. https://doi.org/10.1080/14639220600857639

Gilmore, D. J., & Green, T. R. G. (1984). Comprehension and recall of miniature programs. *International Journal of Man-*

*Machine Studies*, *21*(1), 31–48. https://doi.org/10.1016/S0020-7373(84)80037-1

Ginns, P. (2006). Integrating information: A meta-analysis of the spatial contiguity and temporal contiguity effects. *Learning and Instruction*, *16*(6), 511–525. https://doi.org/10.1016/j.learninstruc.2006.10.001

Gruhn, V., & Laue, R. (2006). Complexity metrics for business process models. In W. Abramowicz & H. C. Mayr (Eds.), *9th International Conference on Business Information Systems, BIS 2006, Proceedings* (Vol. LNI 85, pp. 1–12). Klagenfurt, Austria: Gesellschaft für Informatik (GI).

Hall, R. H., & O'Donnell, A. (1996). Cognitive and affective outcomes of learning from knowledge maps. *Contemporary Educational Psychology*, *21*(1), 94–101. https://doi.org/10.1006/ceps.1996.0008

Haygood, R. C., & Bourne, L. E. (1965). Attribute-and rule-learning aspects of conceptual behavior. *Psychological Review*, *72*(3), 175–195.

Hjalmarsson, A., & Lind, M. (2004). Managing the dynamic agenda in process modelling seminars: Enhancing communication quality in process modelling. In *2nd International Workshop on Action in Language, Organisations and Information Systems, ALOIS 2004, Proceedings* (p. 19). Linköping, Sweden.

Houston, J. P. (1965). Short-term retention of verbal units with equated degrees of learning. *Journal of Experimental Psychology*, *70*(1), 75–78. https://doi.org/10.1037/h0022012

Irani, P., & Ware, C. (2003). Diagramming information structures using 3D perceptual primitives. *ACM Transactions on Computer-Human Interaction*, *10*(1), 1–19. https://doi.org/10.1145/606658.606659

James, W. (1983). *The principles of psychology*. Read Books Ltd. https://doi.org/10.1353/hph.1983.0040

Jensen, K. (1987). Coloured petri nets. In W. Brauer, W. Reisig, & G. Rozenberg (Eds.), *Petri Nets: Central Models and Their Properties* (Vol. LNCS 254). Springer.

Jensen, K. (1991). Coloured Petri nets: A high level language for system design and analysis. In G. Rozenberg (Ed.), *12th International Conference on Application and Theory of Petri Nets, ICATPN 1989, Proceedings* (Vol. LNCS 483, pp. 342–416). Springer. https://doi.org/10.1007/3-540-53863-1_31

Juran, J. M., & Godfrey, A. B. (1998). *Juran's quality handbook*. McGraw-Hill. https://doi.org/10.1108/09684879310045286

Kitchenham, B. A., Brereton, O. P., Budgen, D., Turner, M., Bailey, J., & Linkman, S. (2009). Systematic literature reviews in software engineering: A systematic literature review. *Information and Software Technology*, *51*(1), 7–15. https://doi.org/10.1016/j.infsof.2008.09.009

Koffka, K. (2013). *Principles of Gestalt psychology*. Routledge. https://doi.org/10.2307/2180493

Koshman, S. (2010). Visualizing metadata for environmental datasets. In *10th International Conference on Dublin Core and Metadata Applications, DC 2010, Proceedings* (pp. 11–19). Pittsburgh, Pennsylvania, USA.

Krajcik, J. S. (1991). *The psychology of learning science*. Routledge.

Krogstie, J., Sindre, G., & Jørgensen, H. (2006). Process models representing knowledge for action: A revised quality framework. *European Journal of Information Systems*, *15*(1), 91–102. https://doi.org/10.1057/palgrave.ejis.3000598

Kuhn, M. R., & Stahl, S. A. (2003). Fluency: A review of developmental and remedial practices. *Journal of Educational Psychology*, *95*(1), 3–21. https://doi.org/10.1037/0022-0663.95.1.3

Larkin, J. H., & Simon, H. A. (1987). Why a diagram is (sometimes) worth ten thousand words. *Cognitive Science*, *11*(1), 65–100.

Lindland, O. I., Sindre, G., & Sølvberg, A. (1994). Understanding quality in conceptual modeling. *IEEE Software*, *11*(2), 42–49. https://doi.org/10.1109/52.268955

Linn, M. C., & Muilenburg, L. (1996). Creating lifelong science learners: What models form a firm foundation? *Educational Researcher*, *25*(5), 18–24. https://doi.org/10.3102/0013189X025005018

Lohse, G. L., Min, D., & Olson, J. R. (1995). Cognitive evaluation of system representation diagrams. *Information and Management*, *29*(2), 79–94. https://doi.org/10.1016/0378-7206(95)00024-Q

Long, J. (2002). *Relationships between common graphical representations in system engineering*.

Lorch, R. F. J., Lorch, E. P., & Inman, W. E. (1993). Effects of signaling topic structure on text recall. *Journal of Educational Psychology*. https://doi.org/10.1037//0022-0663.85.2.281

Lowe, R. K. (2003). Animation and learning: Selective processing of information in dynamic graphics. *Learning and Instruction*, *13*(2), 157–176. https://doi.org/10.1016/S0959-47520200018-X

Maiden, N. A., & Sutcliffe, A. G. (1992). Exploiting usable specifiations through analogy. *Communications of the ACM*, *35*(4), 55–64. https://doi.org/10.1145/129852.129857

Mautone, P. D., & Mayer, R. E. (2007). Cognitive aids for guiding graph comprehension. *Journal of Educational Psychology*, *99*(3), 640–652. https://doi.org/10.1037/0022-0663.99.3.640

Mayer, R. E. (1976). Comprehension as affected by structure of problem representation. *Memory & Cognition*, *4*(3), 249–255. https://doi.org/10.3758/BF03213171

Mayer, R. E., & Moreno, R. (2003). Nine ways to reduce cognitive load in multimedia learning. *Journal of Educational Psychology*, *38*(1), 43–52. https://doi.org/10.1207/S15326985EP3801_6

Mayer, R. E., & Sims, V. K. (1994). For whom is a picture worth a thousand words? Extensions of a dual-coding theory of multimedia learning. *Journal of Educational Psychology*, *86*(3), 389–401.

McAndrew. (1983). Underlining and notetaking: Some suggestions from research. *Journal of Reading*, *27*(2), 103–108.

McNamara, D. S., Kintsch, E., Songer, N. B., & Kintsch, W. (1996). Are good texts always better? Interactions of text coherence, background knowledge, and levels of understanding in learning from text. *Cognition and Instruction*, *14*(1), 1–43.

Mendling, J., & Recker, J. (2007). Extending the discussion of model quality: Why clarity and completeness may not always be enough. In B. Pernici & J. A. Gulla (Eds.), *19th International Conference on Advanced Information Systems Engineering, CAiSE 2007, Workshop Proceedings* (Vol. CEUR 365, pp. 101–111). Trondheim, Norway: Tapir Academic Press.

Mendling, J., Reijers, H. A., & Van der Aalst, W. M. P. (2010). Seven process modeling guidelines (7PMG). *Information*

and *Software Technology*, *52*(2), 127–136. https://doi.org/10.1016/j.infsof.2009.08.004

Miller, G. A. (1956). The magical number seven, plus or minus two: Some limits on our capacity for processing information. *Psychological Review*, *63*(2), 81–97. https://doi.org/10.1016/S0895-7177(03)90083-5

Moody, D. L. (1998). Metrics for evaluating the quality of entity relationship models. In T. W. Ling, S. Ram, & M. Li Lee (Eds.), *17th International Conference on Conceptual Modeling, ER 1998, Proceedings* (Vol. LNCS 1507, pp. 211–225). Singapore, China: Springer. https://doi.org/10.1007/978-3-540-49524-6_18

Moody, D. L. (2009). The "physics" of notations: Toward a scientific basis for constructing visual notations in software engineering. *IEEE Transactions on Software Engineering*, *35*(6), 756–779. https://doi.org/10.1109/TSE.2009.67

Moreno, R. (2006). When worked examples don't work: Is cognitive load theory at an impasse? *Learning and Instruction*, *16*(2), 170–181. https://doi.org/10.1016/j.learninstruc.2006.02.006

Moreno, R., & Mayer, R. E. (1999). Cognitive principles of multimedia learning: The role of modality and contiguity. *Journal of Educational Psychology*, *91*(2), 358–368. https://doi.org/10.1037//0022-0663.91.2.358

Morris, E. P. (1901). *On principles and methods in Latin syntax*. C. Scribner's sons.

Neisser, U., & Weene, P. (1962). Hierarchies in concept attainment. *Journal of Experimental Psychology*, *64*(6), 640–645. https://doi.org/10.1037/h0042549

Nelson, H. J., Poels, G., Genero, M., & Piattini, M. (2012). A conceptual modeling quality framework. *Software Quality Journal*, *20*(1), 201–228. https://doi.org/10.1007/s11219-011-9136-9

Newell, A., & Simon, H. A. (1972). *Human problem solving*. Prentice-Hall.

Nonaka, I. (2008). *The Knowledge-Creating Company*. Harvard Business Review Press.

Novak, J. D., & Cañas, A. J. (2008). *The theory underlying concept maps and how to construct and use them. Technical Report IHMC CmapTools*. Pensacola.

Ogao, P. J., & Kraak, M. J. (2002). Defining visualization operations for temporal cartographic animation design. *International Journal of Applied Earth Observation and Geoinformation*, *4*(1), 23–31. https://doi.org/10.1016/S0303-2434(02)00005-3

Ormrod, J. E. (1999). *Human learning* (3rd ed.). Upper Saddle River, New Jersey: Merrill.

Paas, F. G. W. C., Renkl, A., & Sweller, J. (2004). Cognitive load theory: Instructional implications of the interaction between information structures and cognitive architecture. *Instructional Science*, *32*(1/2), 1–8. https://doi.org/10.1023/B:TRUC.0000021806.17516.d0

Paivio, A. (1991). Dual coding theory: Retrospect and current status. *Canadian Journal of Psychology*, *45*(3), 255–287.

Paré, G., Trudel, M. C., Jaana, M., & Kitsiou, S. (2015). Synthesizing information systems knowledge: A typology of literature reviews. *Information and Management*, *52*(2), 183–199. https://doi.org/10.1016/j.im.2014.08.008

Paris, S. G., & Lindauer, B. K. (1976). The role of inference in children's comprehension and memory for sentences. *Cognitive Psychology*, *8*(2), 217–227. https://doi.org/10.1016/0010-0285(76)90024-4

Peebles, D., & Cheng, P. C.-H. (2003). Modeling the effect of task and graphical representation on response latency in a graph reading task. *Human Factors: The Journal of the Human Factors and Ergonomics Society*, *45*(1), 28–46. https://doi.org/10.1518/hfes.45.1.28.27225

Petticrew, M., & Roberts, H. (2006). *Systematic reviews in the social sciences*. Blackwell Publishing. https://doi.org/10.1002/9780470754887

Pilgrim, M. J. (2003). *The application of visualisation techniques to the process of building performance analysis*. Loughborough University.

Pintrich, P. R., & Schunk, D. H. (2002). *Motivation in education: Theory, research and practice*. Merrill.

Pollio, H. R. (1968). *Associative structure and verbal behavior*. Prentice-Hall.

Prezler, R. (2004). Cooperative concept mapping. *Journal of College Science Teaching*, *33*(6).

Purchase, H. C. (2002). Metrics for graph drawing aesthetics. *Journal of Visual Languages & Computing*, *13*(5), 501–516. https://doi.org/10.1016/S1045-926X(02)90232-6

Reber, R., Schwarz, N., & Winkielman, P. (2004). Processing fluency and aesthetic pleasure: Is beauty in the perceiver's processing experience? *Personality and Social Psychology Review*, *8*(4), 364–382. https://doi.org/10.1207/s15327957pspr0804_3

Reber, R., Winkielman, P., & Schwarz, N. (1998). Effects of perceptual fluency on affective judgments. *Psychological Science*, *9*(1), 45–48. https://doi.org/10.1111/1467-9280.00008

Regnell, B., Andersson, M., & Bergstrand, J. (1996). A hierarchical use case model with graphical representation. In *3th IEEE Symposium and Workshop on Engineering of Computer Based Systems, ECBS 1996, Proceedings* (pp. 270–277). Friedrichshafen, Germany: IEEE. https://doi.org/10.1109/ECBS.1996.494538

Reimann, P. (2003). Multimedia learning: Beyond modality. *Learning and Instruction*, *13*(2), 245–252. https://doi.org/10.1016/S0959-47520200024-5

Reynolds, R. E., & Shirey, L. L. (1988). The role of attention in studying and learning. In C. E. Weinstein, E. T. Goetz, & P. A. Alexander (Eds.), *Learning and Study Strategies: Issues in Assessment, Instruction, and Evaluation* (pp. 77–100). Academic Press. https://doi.org/10.1016/B978-0-12-742460-6.50012-8

Roberts, J. C. (2000). Visualization Display models - Ways to classify visual representations. *International Journal of Computer Integrated Design and Construction*, *2*(4), 241–250.

Rockwell, S., & Bajaj, A. (2005). COGEVAL: Applying cognitive theories to evaluate conceptual models. *Advanced Topics in Database Research*, *4*, 255–282. https://doi.org/10.4018/978-1-59140-471-2.ch012

Rogers, Y., & Scaife, M. (1998). How can interactive multimedia facilitate learning? In J. Lee (Ed.), *1st International Workshop On Intelligence and Multimodality in Multimedia Interfaces. Research and Applications. AAAI* (pp. 1–25).

Roth, S. F., & Mattis, J. (1990). Data characterization for intelligent graphics presentation. In *8th Conference on Human Factors in Computing Systems, CHI 1990, Proceedings* (pp. 193–200). Seattle, Washington, USA: ACM. https://doi.org/10.1145/97243.97273

Rowlatt, P. (2008). Consciousness and memory. *Journal of Consciousness Studies*, *16*(December), 68–78.

Sadowska, M. (2013). *Quality of business models expressed in BPMN* (Master thesis). Blekinge Institute of Technology.

Saltz, E. (1971). *The cognitive bases of human learning*. Dorsey Press.

Scaife, M., & Rogers, Y. (1996). External cognition: How do graphical representations work? *International Journal of Human-Computer Studies*, *45*(2), 185–213. https://doi.org/10.1006/ijhc.1996.0048

Schnotz, W. (2013). Integrated model of text and picture comprehension. In R. E. Mayer (Ed.), *The Cambridge Handbook of Multimedia Learning* (pp. 1–49). Cambridge University Press.

Schnotz, W., & Bannert, M. (2003). Construction and interference in learning from multiple representation. *Learning and Instruction*, *13*(2), 141–156. https://doi.org/10.1016/S0959-47520200017-8

Schuette, R., & Rotthowe, T. (1998). The Guidelines of Modeling - An approach to enhance the quality in information models. In T. W. Ling, S. Ram, & M. Li Lee (Eds.), *17th International Conference on Conceptual Modeling, ER 1998, Proceedings* (Vol. LNCS 1507, pp. 240–254). Singapore, China: Springer. https://doi.org/10.1007/978-3-540-49524-6_20

Shannon, C. E., & Weaver, W. (1949). *A mathematical model of communication*. Urbana: University of Illinois Press.

Si-said Cherfi, S., Akoka, J., & Comyn-Wattiau, I. (2002). Conceptual modeling quality - From EER to UML schemas evaluation. In S. Spaccapietra, S. T. March, & Y. Kambayashi (Eds.), *21st International Conference on Conceptual Modeling, ER 2002, Proceedings* (Vol. LNCS 2503, pp. 414–428). Tampere, Finland: Springer. https://doi.org/10.1007/3-540-45816-6_38

Siau, K., & Tan, X. (2005). Improving the quality of conceptual modeling using cognitive mapping techniques. *Data & Knowledge Engineering*, *55*(3), 343–365. https://doi.org/10.1016/j.datak.2004.12.006

Silver, B. (2011). *BPMN: Method & style*. Cody-Cassidy Press.

Smith, W. (2002). A system for monitoring and management of computational grids. In *33rd International Conference on Parallel Processing, ICPP 2002, Proceedings* (pp. 55–62). Vancouver, BC, Canada: IEEE. https://doi.org/10.1109/ICPP.2002.1040859

Snoeck, M., & Poels, G. (2000). Analogical reuse of structural and behavioural aspects of event-based object-oriented domain models. In *11th International Workshop on Database and Expert Systems Applications, DEXA 2000, Proceedings* (pp. 802–806). London, UK: IEEE. https://doi.org/10.1109/DEXA.2000.875117

Spence, R. (2007). *Information visualization: Design for interaction* (2nd ed.). Prentice Hall.

Spires, H. A., & Donley, J. (1998). Prior knowledge activation: Inducing engagement with informational texts. *Journal of Educational Psychology*, *90*(2), 249–260.

Stenning, K., & Oberlander, J. (1995). A cognitive theory of graphical and linguistic reasoning: Logic and implementation. *Cognitive Science*, *19*(1), 97–140. https://doi.org/10.1016/0364-0213(95)90005-5

Sweller, J. (1988). Cognitive load during problem solving: Effects on learning. *Cognitive Science*, *12*(2), 257–285. https://doi.org/10.1016/0364-0213(88)90023-7

Sweller, J. (1994). Cognitive load theory, learning difficulty, and instructional design. *Learning and Instruction*, *4*(4), 295–312. https://doi.org/10.1016/0959-4752(94)90003-5

Sweller, J., Chandler, P., Tierney, P., & Cooper, M. (1990). Cognitive load as a factor in the structuring of technical material. *Journal of Experimental Psychology: General*, *119*(2), 176–192. https://doi.org/10.1037/0096-3445.119.2.176

Sweller, J., Van Merriënboer, J. J. G., & Paas, F. G. W. C. (1998). Cognitive architecture and instructional design. *Educational Psychology Review*, *10*(3), 251–296. https://doi.org/10.1023/A:1022193728205

Tuchman, B. W. (1980, November). The decline of quality. *The New York Times*, p. 38.

Tversky, B. (2001). Spatial schemas in depictions. In *Spatial Schemas and Abstract Thought* (p. 352). Massachusetts: MIT Press.

Tversky, B., Morrison, J. B., & Betrancourt, M. (2002). Animation: Can it facilitate? *International Journal of Human-Computer Studies*, *57*(4), 247–262. https://doi.org/10.1006/ijhc.1017

Van Merriënboer, J. J. G. (1997). *Training complex cognitive skills: A four-component instructional design model for technical training*. Englewood Cliffs, New Jersey: Educational Technology.

Van Merriënboer, J. J. G., Clark, R. E., & Croock, M. B. M. (2002). Blueprints for complex learning: The 4C/ID-model. *Educational Technology Research and Development*, *50*(2), 39–61. https://doi.org/10.1007/BF02504993

Van Merriënboer, J. J. G., & Kester, L. (2005). The four-component instructional design model: Multimedia principles in environments for complex learning. In *The Cambridge Handbook of Multimedia Learning* (pp. 104–148). Cambridge University Press.

Van Merriënboer, J. J. G., & Sweller, J. (2005). Cognitive load theory and complex learning: Recent developments and future directions. *Educational Psychology Review*, *17*(2), 147–177. https://doi.org/10.1007/s10648-005-3951-0

Vanhoucke, M. (2012). *Project management with dynamic scheduling. Baseline scheduling, risk analysis and project control* (2nd ed.). Ghent: Springer. https://doi.org/10.1007/978-3-642-40438-2

Vessey, I., & Galletta, D. F. (1991). Cognitive fit: An empirical study of information acquisition. *Information Systems Research*, *2*(1), 63–84. https://doi.org/10.1287/isre.2.1.63

Viyović, V., Maksimović, M., & Perišić, B. (2014). Sirius: A rapid development of DSM graphical editor. In *18th International IEEE Conference on Intelligent Engineering Systems, INES 2014, Proceedings* (pp. 233–238). Tihany, Hungary: IEEE. https://doi.org/10.1109/INES.2014.6909375

Wagelaar, D., & Van Der Straeten, R. (2007). Platform ontologies for the model-driven architecture. *European Journal of Information Systems*, *16*(4), 362–373. https://doi.org/10.1057/palgrave.ejis.3000686

Wand, Y., & Weber, R. (1990). An ontological model of an information system. *IEEE Transactions on Software Engineering*, *16*(11), 1282–1292. https://doi.org/10.1109/32.60316

Ware, C. (2004). *Information visualization: Perception for design* (2nd ed.). San Francisco: Morgan-Kaufmann.

Ware, C. (2005). Visual queries: The foundation of visual thinking. In S. Tergan & T. Keller (Eds.), *Knowledge Creation*

*Diffusion Utilization* (Vol. LNCS 3426, pp. 27–35). Springer. https://doi.org/10.1007/11510154_2

Ware, C., Purchase, H. C., Colpoys, L., & McGill, M. (2002). Cognitive measurements of graph aesthetics. *Information Visualization*, *1*(2), 103–110. https://doi.org/10.1057/palgrave.ivs.9500013

Webster, J., & Watson, R. T. (2002). Analyzing the past to prepare for the future writing a literature review. *MIS Quarterly*, *26*(2), xiii–xxiii. https://doi.org/10.2307/4132319

Wertheimer, M. (1923). Untersuchungen zur lehre von der Gestalt. II. *Psychologische Forschung*, *4*(1), 301–350. https://doi.org/10.1007/BF00410640

Wheildon, C., & Heard, G. (2005). *Type & layout: Are you communicating or just making pretty shapes?* Worsley Press.

Yeung, A. S., Jin, P., & Sweller, J. (1999). Cognitive load and learner expertise: Split-attention and redundancy effects in reading comprehension tasks with vocabulary definitions. *The Journal of Experimental Education*, *67*(3), 197–217. https://doi.org/10.1080/00220979909598353

Zahid, A., Wanger, L., & Kochevar, P. (1994). An intelligent visualization system for earth science data analysis. *Journal of Visual Languages & Computing*, *5*(4), 307–320. https://doi.org/10.1006/jvlc.1994.1018

Zugal, S. (2013). *Applying cognitive psychology for improving the creation, understanding and maintenance of business process models* (doctoral thesis). Innsbruck University.

Zur Muehlen, M., & Recker, J. (2008). How much language is enough? Theoretical and practical use of the Business Process Modeling Notation. In Z. Bellahsène & M. Léonard (Eds.), *Proc. CAiSE '08* (Vol. LNCS 5074, pp. 465–479). Springer. https://doi.org/10.1007/978-3-540-69534-9_35

# Appendix A. The visual principles of the Gestalt Theory

This appendix presents examples for four exemplary Gestalt Laws.

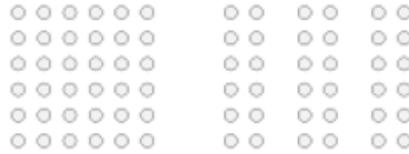

Fig. A.1. Illustration of the Gestalt Law of Proximity. The 72 displayed circles are easily perceived as different groups of circles (left and right group, the right group containing 3 sub-groups)

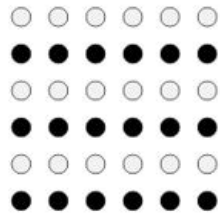

Fig. A.2. Illustration of the Gestalt Law of Similarity. The 36 displayed circles are easily perceived as 6 groups of circles (the combination of color and relative position makes a visual distinction between the 6 'lines' of circles)

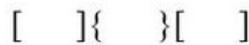

Fig. A.3. Illustration of the Gestalt Law of Symmetry. People are more likely to observe three pairs of brackets rather than six individual ones.

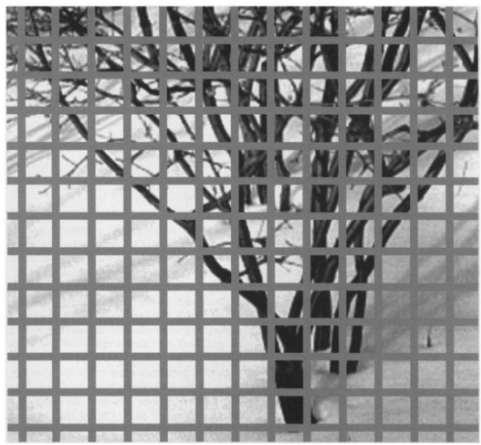

Fig. A.4. Illustration of the Gestalt Law of Good Continuation. The continuity of the branches is easily identified although the grid disrupts the image. Source: Field et al. (1993)

# Appendix B. Text segmentation and labelling versus spatial integration and linking

The pictures below show how, for high-interactive diagrams, spatial integration and linking are less cognitive effective than text segmentation and labelling.

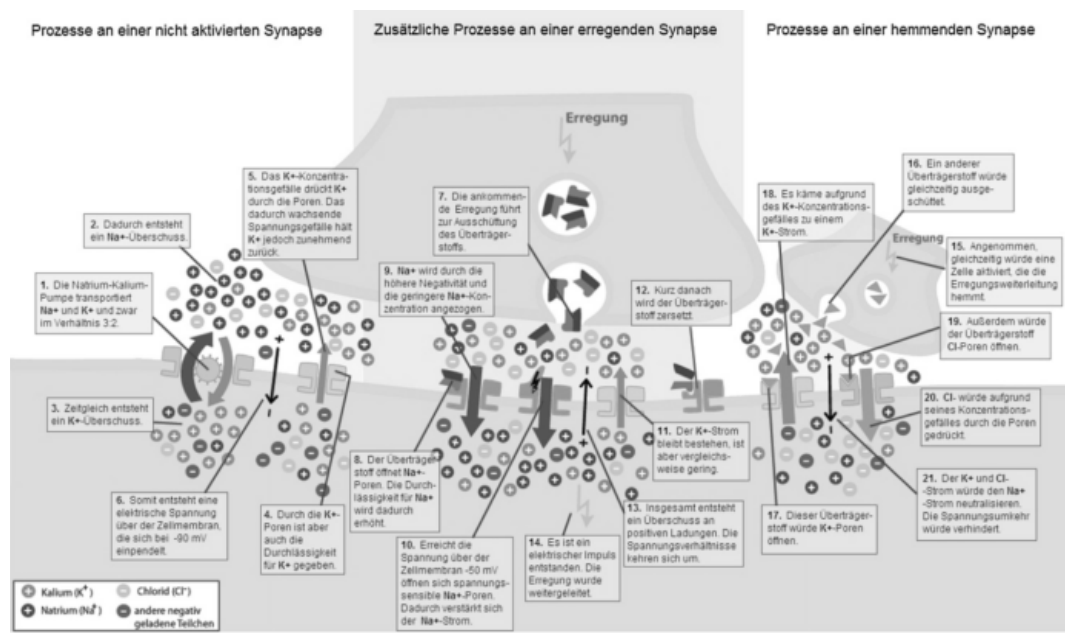

Fig. B.1. Illustration of spatially integrated text and picture. Spatial proximity did not improve learning any further when text was segmented and pictures were labelled. Source: Florax & Ploetzner (2010)

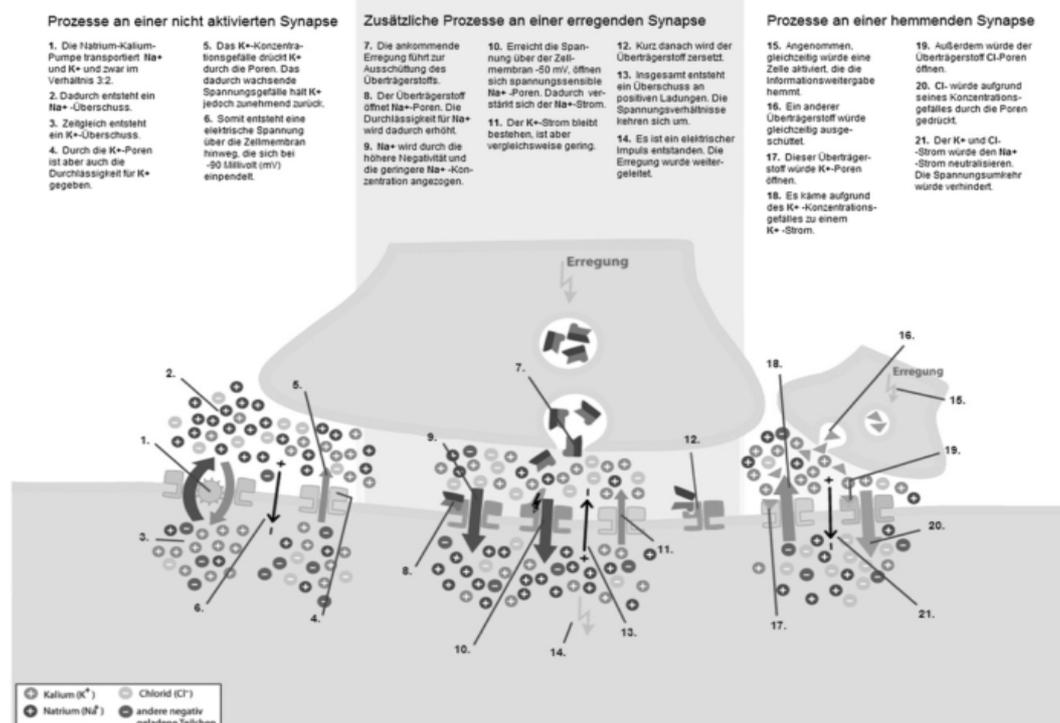

Fig. B.2. Illustration of learning material with segmented text and labelled pictures. Source: Florax & Ploetzner (2010)

# Appendix C. The visual variables of Bertin

Table C.1 presents the eight visual variables defined by Bertin. They are grouped into planar variables, which relate to the positioning of a graphical element, and retinal variables, which relate to the appearance of a graphical element. The *power* indicates which type of values the variable typically represents, and the *capacity* refers to the maximum number of instances of the variable that the diagram should typically not exceed. Fig. C.1 illustrates how the retinal variables can be applied at different levels, i.e., point, line, and area.

Table C.1. The different visual variables: power = scale of measure, capacity = maximal number of instances to be used on a diagram. Source: Moody (2009)

| Type | Variable | Power | Capacity |
| --- | --- | --- | --- |
| Planar variable | Horizontal position | Interval | 10-15 |
|  | Vertical position | Interval | 10-15 |
| Retinal variables | Size | Interval | 20 |
|  | Brightness value | Ordinal | 6-7 |
|  | Color | Nominal | 7-10 |
|  | Texture | Nominal | 2-5 |
|  | Shape | Nominal | 30+ |
|  | Orientation | Nominal | 4 |

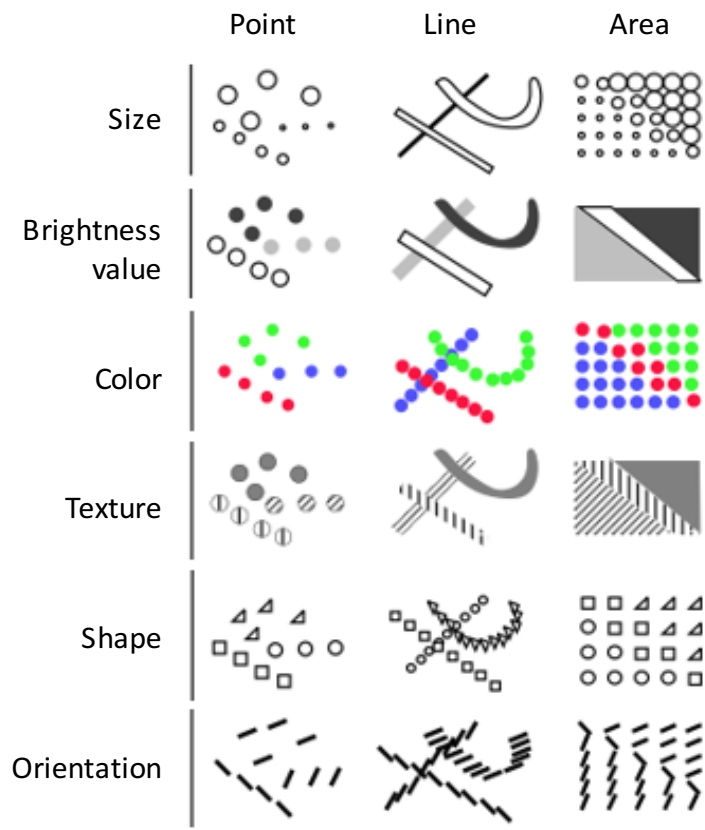

Fig. C.1. The retinal variables apply for points, lines and areas. Source: Roberts (2000)